\documentclass[11pt,floatfix,nofootinbib,showpacs,aps,epsf,amsfonts,amssymb,amsbsy]{revtex4}

\usepackage[T1]{fontenc}
\usepackage[latin1]{inputenc}
\usepackage[english]{babel}
\usepackage{url}
\usepackage{epsfig}
\usepackage{amsmath}
\usepackage{amssymb}
\usepackage{graphics,bm}
\newcommand{\beq}{\begin{equation}}
\newcommand{\eeq}{\end{equation}}
\newcommand{\bqa}{\begin{eqnarray}}
\newcommand{\eqa}{\end{eqnarray}}
\def\sumint{\hbox{$\sum$}\!\!\!\!\!\!\int}
\def\sumintsmall{\hbox{$\sum$}\!\!\!\!\!{\displaystyle \int}}

\begin{document}
\begin{abstract}
We study the thermodynamics of massless $\phi^4$-theory using screened 
perturbation theory. In this method, the perturbative expansion is
reorganized by adding and subtracting a thermal mass term in the Lagrangian.
We calculate the free energy through four loops expanding
in a double power expansion in $m/T$ and $g^2$, where 
$m$ is the thermal mass and 
$g$ is the coupling
constant. The expansion is truncated at order $g^7$ and the 
loop expansion is shown to have better convergence properties than the 
weak-coupling expansion.
The free energy at order $g^6$ involves the four-loop triangle
sum-integral evaluated by Gynther, Laine, Schr\"oder, Torrero, and Vuorinen
using the methods developed by Arnold and Zhai.
The evaluation of the free energy
at order $g^7$ requires the evaluation of a nontrivial three-loop 
sum-integral, which we calculate by the same methods.

\end{abstract}

\title{Four-loop Screened Perturbation Theory}

\author{Jens O. Andersen}
\email{andersen@tf.phys.ntnu.no}
\affiliation{Department of Physics, Norwegian University of Science and 
Technology, N-7491 Trondheim, Norway}
\author{Lars Kyllingstad} 
\email{lars.kyllingstad@ntnu.no}
\affiliation{Department of Physics, Norwegian University of Science and 
Technology, N-7491 Trondheim, Norway}

\date{\today}


\maketitle
\small
\section{Introduction}
In recent years there has been significant progress in the understanding of
thermal field theories in
equilibrium~\cite{francois,dirkie,ulrike,asrev}. 
For example, the thermodynamic functions
can be calculated as power series in the coupling constant $g$ at 
weak coupling and advanced calculational techniques have been developed in 
order to go beyond the first few corrections.  
The pressure has been calculated through order $g^5$ for massless 
$\phi^4$-theory~\cite{parwani1,ericnieto1}, 
massless QED~\cite{coriano,parwani2,joaqed}, and
massless nonabelian gauge theories~\cite{arnold1,zhaikast,ericnieto2}.
Very recently, the calculation frontier has been pushed to order
$g^6$ in massless $\phi^4$-theory by 
Gynther, Laine, Schr\"oder, Torrero, and Vuorinen~\cite{4loopfinns}. 
The calculation in Ref.~\cite{4loopfinns} involves the computation of 
complicated four-loop vacuum diagrams and was motivated by 
the corresponding problem in nonabelian gauge theories:
There are three momentum scales -- hard momenta of order $T$, 
soft momenta of order $gT$, and supersoft momenta of order $g^2T$, which 
give contributions to the free energy.
The contribution from the hard scale $T$ to the free energy can be calculated
as a power series in $g^2$ using na\"ive perturbation theory without 
resummed propagators. The order $g^6$ is the first order at which 
all three momentum scales in QCD  contribute to the
free energy and so it is important to calculate the full $g^6$-term.
Such a calculation involves the evaluation of four-loop vacuum
diagrams in four dimensions.

However, it is well known that the weak-coupling expansion is 
very sensitive to the renormalization scale, and it is furthermore 
convergent only if the coupling constant is tiny.
The physical origin of this instability does not seem to be related to the 
magnetic mass problem in QCD, as it appears in $\phi^4$-theory and QED
as well. Rather, it seems to be associated with screening effects and 
quasiparticles.

In recent years there have been large efforts to reorganize the perturbative
series such that it has improved convergence properties. Several of these
methods are variational in nature, in which the thermodynamic potential
$\Omega$ depends on one or more variational parameters $m_i$.
The pressure and other thermodynamic quantities are then found by
evaluating $\Omega$ and its derivatives at the variational point where
$\delta\Omega/\delta m_i=0$. 

One of these methods is screened perturbation theory (SPT) which in the context
of hot $\phi^4$-theory was introduced by Karsch, Patk\'os and 
Petreczky~\cite{karsch}, see also Refs.~\cite{yuk,steve,janke}.
In this approach, one introduces a single variational parameter $m^2$
which is added to and subtracted from the original Lagrangian.
The added piece is kept as a part of the free Lagrangian and the subtracted
piece is treated as an interaction. The parameter $m^2$ has a simple
interpretation of a thermal mass and satisfies a variational equation.
SPT has been 
applied to calculate the pressure to three-loop order~\cite{abssc} and 
and the convergence properties of the successive approximations are
dramatically improved as compared to the weak-coupling expansion.
The mass parameter is of order $g$ and so it might be reasonable to
carry out an additional expansion of the Feynman diagrams in powers of
$m/T$, and truncate at the appropriate order. This was done in Ref.~\cite{asmot}
and it was demonstrated that the double expansion in $m/T$ and $g$
converges quickly to the numerically exact result even for large values
of the coupling.

The generalization of SPT to gauge theories cannot simply be made 
by adding and subtracting a local mass term as this would violate
gauge invariance. Instead one adds and subtracts to the Lagrangian
a hard thermal loop (HTL) improvement term~\cite{bphtl}.
The free piece of the Lagrangian includes the HTL self-energies,
while the remaining terms are treated as perturbations.
Hard-thermal loop perturbation theory is a manifestly gauge invariant approach
that can be used to calculate static as well as dynamic quantities
in a systematic expansion. HTL perturbation theory has been
applied to calculate the pressure to 
two-loop order~\cite{absprl,absgl1,absq1,absgl2,absq2} in an $m/T$
expansion 
and the convergence properties of the successive approximations are again
improved as compared to the weak-coupling expansion.

Another variational method in which the propagator is a variational function
was constructed by Luttinger and Ward~\cite{ward} and by Baym~\cite{baym}
for nonrelativistic fermions
in the early 1960s. Later, it was generalized to relativistic quantum
field theories by Cornwall, Jackiw and Tomboulis~\cite{cjt}.
The approach is based on the fact that the thermodynamic potential
$\Omega$ can be written in terms of the two-particle irreducible (2PI)
vacuum diagrams. The propagator $D$ satisfies the 
variational equation $\delta\Omega/\delta D=0$.
The 2PI effective action formalism is also referred to as $\Phi$-derivable
approximations.


Since the 2PI effective action formalism involves an effective 
propagator, a truncated calculation in the loop expansion or
$1/N$-expansion involves a selective resummation of diagrams from all
orders of perturbation theory. This fact makes renormalization of 
$\Phi$-derivable approximations highly nontrivial.
In recent years, there have been large efforts to prove renormalizability
in the loop expansion, $1/N$-expansion, 
or the Hartree approximation, 
and in particular
to prove that the counterterms are medium independent, i.e. independent
of temperature and chemical potential~\cite{knoll,urko,b2,fejos}.

The second issue is that of gauge-fixing dependence.
While the exact 2PI effective action is gauge independent
at the stationary point, this property is often lost in 
approximations. The problem has been examined
by Arrizabalaga and Smit~\cite{arri} as well as 
Carrington {\it et al}~\cite{meg}. In Ref.~\cite{arri}, it 
was shown that the $n$-loop 
$\Phi$-derivable approximation, which is defined by
the truncation of the action functional after $n$ loops, has
a gauge dependence that shows up at order $g^{2n}$. 
Furthermore, if the $n$th order solution to the gap equation is
used to evaluate the complete effective action, the gauge
dependence first shows up at order $g^{4n}$. Explicit examples
of the gauge dependence of the three-loop $\Phi$-derivable
approximation can be found in Ref.~\cite{jmqed}.

The $\Phi$-derivable approach has been used by
Blaizot, Iancu, and Rebhan~\cite{bir1,bir2,bir3} and by Peshier~\cite{peshier}
to calculate the thermodynamic quantities at the two-loop level
in scalar field theory as well as in gauge theories.
The calculations are based on the fact that the solution to the 
gap equation for the propagator for soft momenta is given by the HTL
self-energies. Three-loop calculations have been performed 
in scalar field theory by Braaten and
Petitgirard~\cite{BP-01}, and in QED in Ref.~\cite{jmqed} using an $m/T$
expansion similar to that employed in SPT in Ref.~\cite{asmot}.
The convergence of the successive approximations to the pressure
is improved significantly
compared to the weak-coupling expansion and the sensitivity to the
renormalization scale is also reduced.
In Ref.~\cite{bergesscalar}, the authors carried out a numerically exact
three-loop calculation of the pressure in $\phi^4$-theory.
Similarly, numerically exact two-loop calculations of the pressure in QED
including an analysis of the gauge dependence 
of the results can be found in Ref.~\cite{borsa}.
In these calculations no attempts to compare with the $m/T$ expansions 
of Refs.~\cite{BP-01,jmqed} were made.

Finally, we mention other related resummation methods that have been 
applied in recent years, namely, the 2-particle point irreducable (2PPI) method~\cite{verschelde,ver2} 
as well as the linear delta-expansion~\cite{ram1,ram2,ram3,hf}.
These methods are also variational in spirit. Moreover, it has been shown 
that they correctly  predict a second-order phase transition when applied
to $\phi^4$-theory. In the case of the linear delta-expansion, the 
successive approximations of e.g. the pressure are remarkably stable 
as compared to the weak-coupling expansion.

The article is organized as follows. In Sec.~II, we briefly discuss the 
systematics of screened perturbation theory. In Sec.~III, we calculate
the pressure to four-loop order in a double expansion in $m/T$ and $g^2$.
In Sec.~IV, we discuss different gap equations that are used to determine
the mass parameter in screened perturbation theory. We also
present our numerical results and compare them with the weak-coupling
expansion. In Sec.~V, we summarize. In Appendix A and B, we list the
sum-integrals and the integrals that we need. In Appendix C, we 
discuss the $m/T$ expansion of typical sum-integrals that appear in the
calculation. In Appendix D, we calculate explicitly a
new three-loop sum-integral that contribute to order $g^7$ in 
the $m/T$ expansion.

\section{Screened Perturbation Theory}
The Lagrangian density for a massless scalar field with a $\phi^4$ interaction 
is
\bqa
\label{ori}
{\cal L}={1\over2}\partial_{\mu}\phi\partial^{\mu}\phi
-{g^2\over24}\phi^4+\Delta{\cal L}\;,
\eqa
where $g$ is the coupling constant and $\Delta{\cal L}$
includes counterterms.
Renormalizability guarantees that $\Delta{\cal L}$ is of the form
\bqa
\Delta{\cal L} = {1\over2} \Delta Z \, \partial_{\mu}\phi\partial^{\mu}\phi
	- {1\over24} \Delta g^2 \phi^4 \; .
\eqa

Screened perturbation theory, which was introduced 
in thermal field theory by Karsch, Patk\'os
and Petreczky~\cite{karsch}, is simply a reorganization of the perturbation
series for thermal field theory.
It can be made more systematic by using a framework called 
``optimized perturbation theory'' that Chiku and Hatsuda~\cite{Chiku-Hatsuda}
have applied to a spontaneously broken scalar field theory. 
The Lagrangian density is written as 
\bqa
\label{SPT}
{\cal L}_{\rm SPT}=-{\cal E}_0+{1\over2}\partial_{\mu}\phi\partial^{\mu}\phi
-{1\over2}\left(m^2-m_1^2\right)\phi^2
-{g^2\over24}\phi^4
+\Delta{\cal L}+\Delta{\cal L}_{\rm SPT}
\;.
\eqa
Here, ${\cal E}_0$ is the vacuum energy density term, and we have added and
subtracted mass terms. If we set ${\cal E}_0=0$ and $m_1^2=m^2$, we recover
the original Lagrangian Eq.~(\ref{ori}).
Screened perturbation theory is defined by taking $m^2$ to be of order unity
and $m_1^2$ to of order $g^2$, expanding systematically in powers of
$g^2$ and setting $m_1^2=m^2$ at the end of the calculation.
This defines a reorganization of the perturbative series in which the 
expansion is about the free field theory defined by
\bqa
{\cal L}_{\rm free}=
-{\cal E}_0+{1\over2}\partial_{\mu}\phi\partial^{\mu}\phi
-{1\over2}m^2\phi^2\;.
\eqa
The interacting term is 
\bqa
{\cal L}_{\rm int}=
{1\over2}m_1^2\phi^2
-{g^2\over24}\phi^4
+\Delta{\cal L}+\Delta{\cal L}_{\rm SPT}\;.
\eqa
Screened perturbation theory generates new ultraviolet divergences, but they
can be cancelled by the additional counterterm in $\Delta{\cal L}_{\rm SPT}$.
If we use dimensional regularization and minimal subtraction, the coefficients
of these operators are polynomials in $g^2$ and $(m^2-m_1^2)$.
The counterterm $\Delta {\cal L}$ is
\bqa
\Delta{\cal L}&=&-{\Delta g^2\over24}\phi^4\,.
\eqa
The additional counterterms required to remove the new divergences are
\bqa
\Delta{\cal L}_{\rm SPT}=-\Delta{\cal E}_0-{1\over2}\left(
\Delta m^2-\Delta m_1^2
\right)\phi^2\;.
\eqa
Several terms in the power series expansions of the
counterterms are known from previous calculations at zero temperature.
The counterterms $\Delta g^2$ and $\Delta m^2$ are known to order 
$\alpha^5$, where $\alpha=g^2/(4\pi)^2$~\cite{Kleinert}. 
We will need the coupling constant counterterm to next-to-leading
order in $\alpha$:
\bqa
\Delta g^2=\left[
{3\over2\epsilon}\alpha+\left({9\over4\epsilon^2}-{17\over12\epsilon}\right)
\alpha^2+\cdots
\right]g^2\;.
\eqa
We need the mass counterterms $\Delta m^2$ and $\Delta m^2_1$
to next-to-leading in $\alpha$:
\bqa
\label{dmm}
\Delta m^2&=&\left[{1\over2\epsilon}\alpha+\left({1\over2\epsilon^2}
-{5\over24\epsilon}\right)\alpha^2+\cdots
\right]m^2\;, \\
\label{d1m1}
\Delta m^2_1&=&\left[{1\over2\epsilon}\alpha+\left({1\over2\epsilon^2}
-{5\over24\epsilon}\right)\alpha^2+\cdots
\right]m^2_1\;.
\eqa
The counterterm for $\Delta{\cal E}_0$
has been calculated to order $\alpha^4$~\cite{kastening}.
We will need its expansion only to first order 
in $\alpha$ and second order in $m_1^2$:
\bqa
(4\pi)^2\Delta{\cal E}_0&=&\left[
{1\over4\epsilon}+{1\over8\epsilon^2}\alpha
\right]m^4
\label{de}
-2\left[{1\over4\epsilon}+{1\over8\epsilon^2}\alpha
\right]m_1^2m^2
+\left[{1\over4\epsilon}+{1\over8\epsilon^2}\alpha
\right]m_1^4\;.
\eqa

\section{Free energy to four loops}

In this section, we calculate the $m/T$ expansions of the 
pressure
to four loops in screened perturbation theory.  In performing
the truncation, $m$ is treated as a quantity that is ${\cal O}(g)$
and we include all terms which contribute to order $g^7$.


\subsection{One-loop free energy}
The free energy at leading order in $g^2$ is
\bqa
{\cal F}_0&=&{\cal E}_0+{\cal F}_{\rm 0a}+\Delta_0{\cal E}_0\;,
\eqa
where $\Delta_0{\cal E}_0$ is the term of order $g^0$ in the vacuum energy
counterterm Eq.~(\ref{de}). 
\begin{figure}[htb]
    \center
    \includegraphics{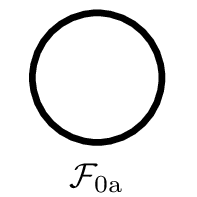}
    \caption{One-loop vacuum diagram.}
\label{diagramf0}
\end{figure}

The expression for diagram ${\cal F}_{\rm 0a}$ in Fig.~\ref{diagramf0} is 
\bqa
{\cal F}_{\rm 0a}&=&{1\over2}\sumint_{P}\log\left[P^2+m^2\right]\;,
\label{f000}
\eqa
where the symbol $\sumintsmall_P$ is defined in Appendix~\ref{appa}.

Treating $m$ as ${\cal O}(gT)$ and including all terms which contribute
through ${\cal O}(g^7)$, we obtain
\bqa
{\cal F}_{\rm 0a}&=&
{1\over2}{\cal I}_0^{\prime}
+{1\over2}m^2{\cal I}_1
+{1\over2}TI_0^{\prime}
-{1\over4}m^4{\cal I}_2
+{1\over6}m^6{\cal I}_3
\;,
\eqa
where the sum-integrals ${\cal I}^\prime_0$ and
${\cal I}_n$ are defined in
Appendix \ref{appa} and the integral $I^\prime_0$ is defined in Appendix
\ref{appb}. In Appendix C, we illustrate the $m/T$ expansion of simple
one-loop sum-integrals such as the one appearing in Eq.~(\ref{f000}).
We also note that most of the multiloop diagrams are products
of simple one-loop sum-integrals. 

The term ${\cal I}_2$ is logarithmically divergent and the 
pole in $\epsilon$ is cancelled by the zeroth-order term 
$\Delta_0{\cal E}_0$ in Eq.~(\ref{de}).
The final result for the truncated one-loop free energy is
\bqa
{\cal F}_0&=&
-{\pi^2T^4\over90}
\Bigg[1-15\hat{m}^2+60\hat{m}^3+45(L+\gamma_E)\hat{m}^4
-{15\over2}\zeta(3)\hat{m}^6
\Bigg]\;,
\label{f0}
\eqa
where $\hat{m}={m\over2\pi T}$ and $L=\log{\mu\over4\pi T}$.

\subsection{Two-loop free energy}
The contribution to the free energy at two loops is given by
\bqa
\label{f1}
{\cal F}_1&=&{\cal F}_{\rm 1a}
+{\cal F}_{\rm 1b}
+\Delta_1{\cal E}_0
+{\partial{\cal F}_{\rm 0a}\over\partial m^2}\Delta_1m^2\;,
\eqa
where $\Delta_1{\cal E}_0$ and $\Delta_1m^2$ are the 
vacuum and mass counterterms of order $g^2$, respectively.
\begin{figure}[htb]
    \center
    \includegraphics{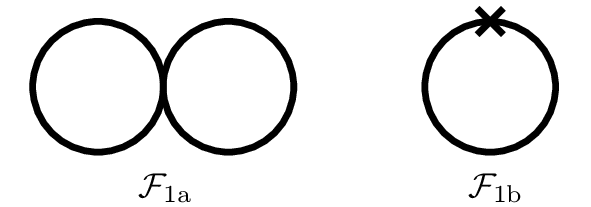}
    \caption{Two-loop vacuum diagrams. The cross denotes a mass insertion.}
\label{diagramf1}
\end{figure}
The expressions for the diagrams 
${\cal F}_{\rm 1a}$ and ${\cal F}_{\rm 1b}$ in Fig.~\ref{diagramf1} are
\bqa
\label{1a0}
{\cal F}_{\rm 1a}&=&{1\over8}g^2\left(\sumint_{P}{1\over P^2+m^2}\right)^2\;,
\\
\label{1b0}
{\cal F}_{\rm 1b}&=&-{1\over2}m_1^2\sumint_{P}{1\over P^2+m^2}\;.
\eqa
Expanding the sum-integrals in Eqs.~(\ref{1a0}) and~(\ref{1b0}) to order
${\cal O}(g^7)$ yields
\bqa
{\cal F}_{\rm 1a}&=&{1\over8}g^2
\Bigg[
{\cal I}_1^2
+2TI_1{\cal I}_1
-2m^2{\cal I}_1{\cal I}_2+T^2I_1^2
-2m^2I_1T{\cal I}_2
+2m^4{\cal I}_1{\cal I}_3+m^4{\cal I}_2^2
+2m^4TI_1{\cal I}_3
\Bigg]\;,
\label{f1a}
\\
\label{f1b}
{\cal F}_{\rm 1b}&=&
-{1\over2}m_1^2\Bigg[
{\cal I}_1+TI_1-m^2{\cal I}_2
+m^4{\cal I}_3
\Bigg]\;,
\eqa
where the integral $I_n$ is defined in Appendix \ref{appb}.

The poles in $\epsilon$ in Eqs.~(\ref{f1a}) 
and~(\ref{f1b}) are cancelled by the counterterms
in Eq.~(\ref{f1}). 
The final result for the two-loop contribution to the free energy is
\bqa\nonumber
{\cal F}_1&=&
{\pi^2T^4\over90}\alpha
\Bigg[
{5\over4}
-15\hat{m}
-15\left(
L+\gamma_E-3\right)\hat{m}^2
+90(L+\gamma_E)\hat{m}^3
+45\left(\left(L+\gamma_E\right)^2+{1\over12}\zeta(3)\right)\hat{m}^4
-{45\over2}\zeta(3)\hat{m}^5
\Bigg]
\\ &&
-{\pi^2T^4\over90}15\hat{m}_1^2
\Bigg[1-6\hat{m}-6
(L+\gamma_E)\hat{m}^2+{3\over2}\zeta(3)\hat{m}^4
\Bigg]\;.
\label{2free}
\eqa
Note that we here and in the following have pulled out a factor of 
${\cal F}_{\rm ideal }=-\pi^2T^4/90$.

\subsection{Three-loop free energy}
The contribution to the free energy at three loops is
\bqa\nonumber
{\cal F}_2&=&{\cal F}_{\rm 2a}+{\cal F}_{\rm 2b}
+{\cal F}_{\rm 2c}+{\cal F}_{\rm 2d}+\Delta_2{\cal E}_0
+{\partial{\cal F}_{\rm 0a}\over\partial m^2}\Delta_2m^2
+{1\over2}{\partial^2{\cal F}_{\rm 0a}\over\left(\partial m^2\right)^2}
\left(\Delta_1m^2\right)^2
+\left({\partial{\cal F}_{\rm 1a}\over\partial m^2}
+{\partial{\cal F}_{\rm 1b}\over\partial m^2}
\right)\Delta_1m^2
\\ &&
+{{\cal F}_{\rm 1a}\over g^2}\Delta_1g^2
+{{\cal F}_{\rm 1b}\over m_1^2}\Delta_1m_1^2\;,
\label{f2}
\eqa
where we have included all necessary counterterms.
\begin{figure}[htb]
    \center
    \includegraphics{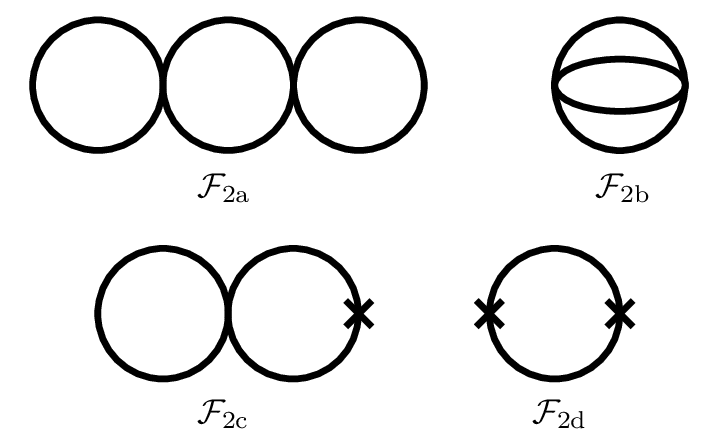}
    \caption{Three-loop vacuum diagrams.}
\label{diagramf3}
\end{figure}
The expressions for the diagrams ${\cal F}_{\rm 2a}$, ${\cal F}_{\rm 2b}$, 
${\cal F}_{\rm 2c}$, and ${\cal F}_{\rm 2d}$
in Fig.~\ref{diagramf3} are
\bqa
{\cal F}_{\rm 2a}&=&
-{1\over16}g^4\left(\sumint_{P}{1\over P^2+m^2}\right)^2
\sumint_{Q}{1\over(Q^2+m^2)^2}\;, \\
\label{massen}
{\cal F}_{\rm 2b}&=&
-{1\over48}g^4\sumint_{PQR}
{1\over P^2+m^2}{1\over Q^2+m^2}{1\over R^2+m^2}{1\over (P+Q+R)^2+m^2}
\;,\\
{\cal F}_{\rm 2c}&=& 
{1\over4}g^2m_1^2\sumint_{P}{1\over P^2+m^2}
\sumint_{Q}{1\over(Q^2+m^2)^2}
\;,\\
{\cal F}_{\rm 2d}&=&
-{1\over4}m_1^4\sumint_{P}{1\over(P^2+m^2)^2}
\;.
\eqa
Expanding in powers of $m^2$ to the appropriate order gives~\footnote{Notice
that the term $TI_1{\cal I}_{\rm sun}$ in ${\cal F}_{\rm 2b}$ in Eq.\ eqref{2b} vanishes. However,
we include this term
because it gives rise to a finite term at four loops
when renormalizing the coupling constant $g$.}
\bqa\nonumber
{\cal F}_{\rm 2a}&=& 
-{1\over16}g^4\Bigg[
T{\cal I}_1^2I_2
+{\cal I}_1^2{\cal I}_2+2T^2{\cal I}_1I_1I_2
+T^3I_1^2I_2
+2TI_1{\cal I}_1{\cal I}_2
\label{2a}
-2m^2T{\cal I}_1{\cal I}_2I_2
+T^2{\cal I}_2I_1^2
-2m^2{\cal I}_1^2{\cal I}_3
\\&&
-2m^2T^2{\cal I}_2I_1I_2
-2m^2{\cal I}_1{\cal I}_2^2
-4m^2TI_1{\cal I}_1{\cal I}_3-2m^2TI_1{\cal I}_2^2
+m^4TI_2{\cal I}_2^2
+2m^4TI_2{\cal I}_1{\cal I}_3
\Bigg]\;,\\
\label{2b}
{\cal F}_{\rm 2b}&=& 
-{1\over48}g^4\Bigg[{\cal I}_{\rm ball}+
T^3I_{\rm ball}+4T I_1 {\cal I}_{\rm sun}
+6T^2{\cal I}_2I_1^2-4m^2{\cal I}_{\rm ball}^{\prime}
-8m^2TI_1\sumint_{QR}{Q^2+(2/d){\bf q}^2\over Q^6R^2(Q+R)^2}
\Bigg]
\;,\\
\label{2c}
{\cal F}_{\rm 2c}&=& 
{1\over4}g^2m_1^2\Bigg[
T{\cal I}_1I_2+{\cal I}_1{\cal I}_2
+T^2I_1I_2+T{\cal I}_2I_1-m^2T{\cal I}_2I_2
-m^2{\cal I}_2^2-2m^2{\cal I}_1{\cal I}_3
-2m^2TI_1{\cal I}_3+m^4TI_2{\cal I}_3
\Bigg]
\;,\\
\label{2d}
{\cal F}_{\rm 2d}&=& 
-{1\over4}m_1^4\Bigg[
TI_2+{\cal I}_2
-2m^2{\cal I}_3
\Bigg]\;,
\eqa
where ${\cal I}_{\rm sun}$, and ${\cal I}_{\rm ball}$, and
${\cal I}_{\rm ball}^{\prime}$ are defined in Appendix~\ref{appa}, and
$I_{\rm ball}$ is defined in Appendix~\ref{appb}.

The poles in $\epsilon$ in Eqs.~(\ref{2a})--(\ref{2d}) 
are cancelled by the counterterms
in Eq.~(\ref{f2}).

The final result for the three-loop contribution to the free energy is
\bqa\nonumber
{\cal F}_2&=&
-{\pi^2T^4\over90}{5\over8}{1\over\hat{m}}\alpha^2\Bigg[
1-2
\Bigg(
{59\over15}-\gamma_E-3L-4{\zeta^{\prime}(-1)\over\zeta(-1)}
+2{\zeta^{\prime}(-3)\over\zeta(-3)}
\Bigg)\hat{m}
\\\nonumber
&& 
-12\hat{m}^2
\Bigg(
5+7L+3\gamma_E
-8\log\hat{m}-8\log2
-4{\zeta^{\prime}(-1)\over\zeta(-1)}
\Bigg)
+
       \bigg(
            268 (L + \gamma_E) - 48 (L + \gamma_E)^2
            + \frac{\zeta'(-1)}{\zeta(-1)} (34 + 12 \gamma_E)
    \nonumber \\    &&
            + 12 \frac{\zeta''(-1)}{\zeta(-1)}
            + \gamma_E (17 - 21 \gamma_E)
            + 34 + \frac{9 \pi^2}{2} - 48 \gamma_1
            - \zeta(3)
            - 6 C_{\rm ball}^{\prime}
        \bigg) \hat m^3
    \nonumber \\ &&
        + 
\bigg(89 + 120 (L+\gamma_E) + [18 (L+\gamma_E)]^2 + 15 \zeta(3) \bigg) \hat m^4
\Bigg]
\nonumber \\
&&
+{\pi^2T^4\over90}{15\over2}{\hat{m}_1^2\over\hat{m}}\alpha\Bigg[
1+2
\left(L+\gamma_E-3\right)\hat{m}
-18(L+\gamma_E)\hat{m}^2
-\left(12\left(L+\gamma_E\right)^2+\zeta(3)\right)\hat{m}^3
+{15\over2}\zeta(3)\hat{m}^4
\Bigg]
\nonumber \\ &&
-{\pi^2T^4\over90}{45\over2}{\hat{m}_1^4\over\hat{m}}
\Bigg[1+2
(L+\gamma_E)\hat{m}-\zeta(3)\hat{m}^3\Bigg]
\;.
\label{3free}
\eqa
Here $C_{\rm ball}^{\prime}=48.7976$ 
is the numerical constant in 
$\mathcal I_\mathrm{ball}^{\prime}$~\cite{4loopfinns}.

\subsection{Four-loop free energy}
The contributions to the free energy at four loops are
\bqa\nonumber
{\cal F}_3&=&{\cal F}_{\rm 3a}+{\cal F}_{\rm 3b}+{\cal F}_{\rm 3c}+
{\cal F}_{\rm 3d}
+{\cal F}_{\rm 3e}+{\cal F}_{\rm 3f}+{\cal F}_{\rm 3g}+{\cal F}_{\rm 3h}
+{\cal F}_{\rm 3i}+{\cal F}_{\rm 3j}
+\Delta_3{\cal E}_0+{\partial {\cal F}_{\rm 0a}\over\partial m^2}\Delta_3m^2
+{1\over6}{\partial^3{\cal F}_{\rm 0a}\over(\partial m^2)^3}
\left(\Delta_1m^2\right)^3
\\ &&\nonumber
+{\partial^2{\cal F}_{\rm 0a}\over(\partial m^2)^2}
\left(\Delta_1m^2\right)\left(\Delta_2m^2\right)
+\left({\partial{\cal F}_{\rm 1a}\over\partial m^2}
+{\partial{\cal F}_{\rm 1b}\over\partial m^2}\right)\Delta_2m^2
+{{\cal F}_{\rm 1a}\over g^2}\Delta_2g^2
+\left(2{{\cal F}_{\rm 2a}\over g^2}
+2{{\cal F}_{\rm 2b}\over g^2}
+{{\cal F}_{\rm 2c}\over g^2}
\right)\Delta_1g^2
\\ && \nonumber
+{1\over2}\left({\partial{\cal F}_{\rm 1a}^2\over(\partial m^2)^2}
+{\partial{\cal F}_{\rm 1b}^2\over(\partial m^2)^2}
\right)(\Delta_1m^2)^2+{{\cal F}_{1b}\over m_1^2}\Delta_2m_1^2
+{\;\;\partial{\cal F}_{\rm 1b}\over m_1^2\partial m^2}
\left(\Delta_1m^2\right)\left(\Delta_1m_1^2\right)
+{1\over g^2}{\partial{\cal F}_{\rm 1a}\over\partial m^2}
\left(\Delta_1g^2\right)
\left(\Delta_1m^2\right)
\\ &&
+\left({{\cal F}_{2c}\over m_1^2}
+2{{\cal F}_{2d}\over m_1^2}\right)\Delta_1m^2_1
+\left({\partial{\cal F}_{\rm 2a}\over\partial m^2}
+{\partial{\cal F}_{\rm 2b}\over\partial m^2}
+{\partial{\cal F}_{\rm 2c}\over\partial m^2}
+{\partial{\cal F}_{\rm 2d}\over\partial m^2}
\right)\Delta_1m^2
\;.
\label{fireloop}
\eqa
Note that some of the terms first contribute at order $g^8$ or higher.
For example, the vacuum counterterm $\Delta_3{\cal E}_0$ first contributes
at order $m^4\alpha^2\sim g^8$.

\begin{figure}[htb]
\begin{center}
    \includegraphics{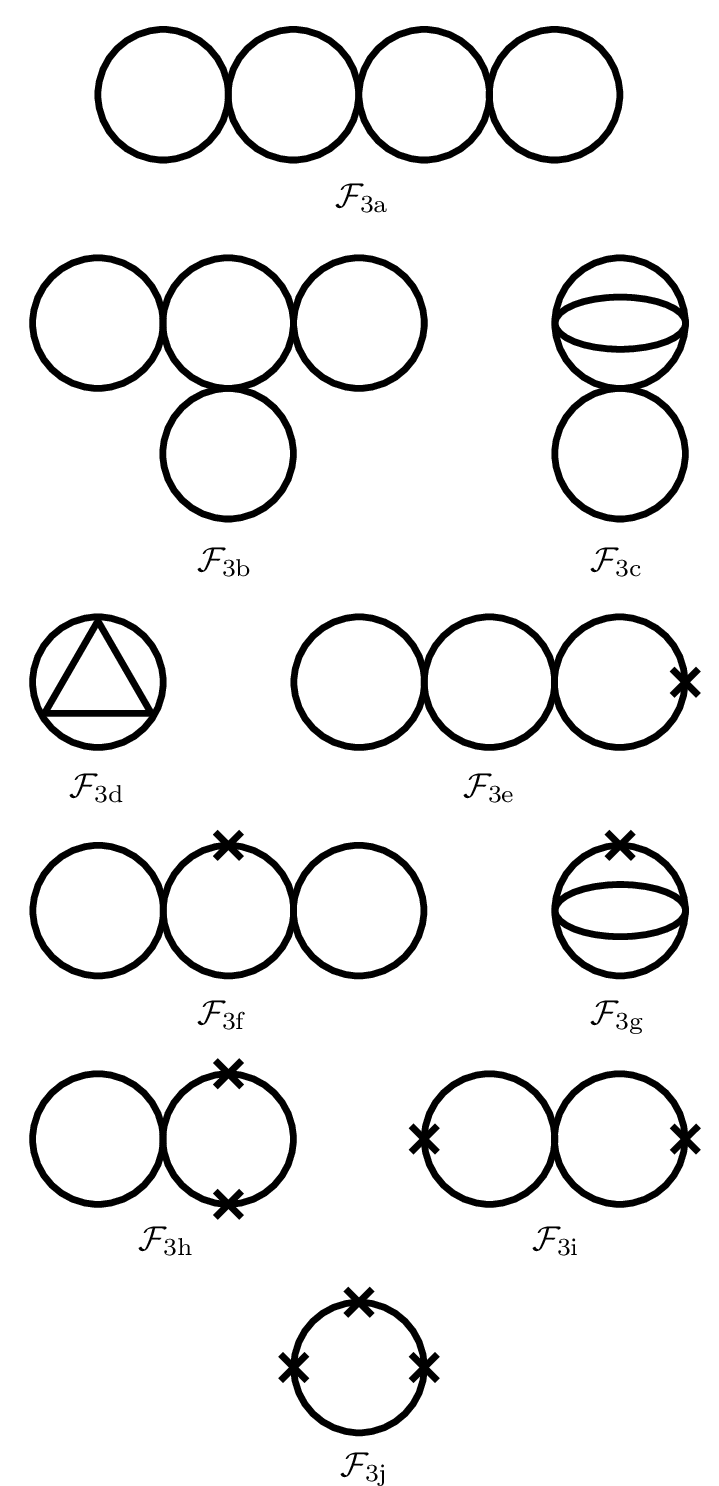}
    \caption{Four-loop vacuum diagrams.}
\label{diagramf4}
\end{center}
\end{figure}

The expressions for the diagrams 
${\cal F}_{\rm 3a}$--${\cal F}_{\rm 3j}$, 
in Fig.~\ref{diagramf4} are
\bqa
{\cal F}_{\rm 3a}&=&
{1\over32}g^6\left(\sumint_{P}{1\over P^2+m^2}\right)^2
\left(\sumint_{Q}{1\over(Q^2+m^2)^2}\right)^2\;, \\
{\cal F}_{\rm 3b}&=&{1\over48}g^6\left(\sumint_P{1\over P^2+m^2}\right)^3
\sumint_Q{1\over(Q^2+m^2)^3}\;,\\
{\cal F}_{\rm 3c}&=&{1\over24}g^6
\sumint_{PQR}
{1\over(P^2+m^2)^2}{1\over Q^2+m^2}{1\over R^2+m^2}{1\over (P+Q+R)^2+m^2}
\sumint_S{1\over S^2+m^2}
\\
{\cal F}_{\rm 3d}&=&{1\over48}g^6
\sumint_{PQRS}{1\over Q^2+m^2}{1\over(P+Q)^2+m^2}
{1\over R^2+m^2}{1\over(P+R)^2+m^2}
{1\over S^2+m^2}{1\over(P+S)^2+m^2}\;, \\
{\cal F}_{\rm 3e}&=&-{1\over8}g^4m_1^2
\sumint_{P}{1\over P^2+m^2}\left(\sumint_{Q}{1\over(Q^2+m^2)^2}\right)^2
\;,\\{\cal F}_{\rm 3f}&=&-{1\over8}g^4m_1^2
\left(\sumint_{P}{1\over P^2+m^2}\right)^2\sumint_{Q}{1\over(Q^2+m^2)^3}
\;,\\
{\cal F}_{\rm 3g}&=&-{1\over12}g^4m_1^2\sumint_{PQR}
{1\over(P^2+m^2)^2}{1\over Q^2+m^2}{1\over R^2+m^2}{1\over (P+Q+R)^2+m^2} \\
{\cal F}_{\rm 3h}&=&{1\over4}g^2m_1^4
\sumint_P{1\over P^2+m^2}
\sumint_Q{1\over(Q^2+m^2)^3}\;,
\\
{\cal F}_{\rm 3i}&=&{1\over8}g^2m_1^4
\left(\sumint_P{1\over(P^2+m^2)^2}\right)^2\;,
\\
{\cal F}_{\rm 3j}&=&-{1\over6}m_1^6\sumint_P{1\over(P^2+m^2)^3}\;.
\eqa

Expanding the sum-integrals in powers of $m^2$ to the appropriate order
gives 
\bqa\nonumber
{\cal F}_{\rm 3a}&=&{1\over32}g^6\left[
T^2I_2^2{\cal I}_1^2+2T^3I_1I_2^2{\cal I}_1
+2TI_2{\cal I}_1^2{\cal I}_2
+{\cal I}_1^2{\cal I}_2^2
+T^4I_1^2I_2^2
+2TI_1{\cal I}_1{\cal I}_2^2
-2m^2T^2I_2^2{\cal I}_1{\cal I}_2
\right.\\&&\left.
-2m^2T^3I_1I_2^2{\cal I}_2
+4T^2I_1I_2{\cal I}_1{\cal I}_2
-4m^2TI_2{\cal I}_1{\cal I}_2^2+2T^3I_1^2I_2{\cal I}_2
-4m^2TI_2{\cal I}_1^2{\cal I}_3
\label{first4}
\right]\;,\\ \nonumber
{\cal F}_{\rm 3b}&=&{1\over48}g^6\left[
TI_3{\cal I}_1^3+3T^2I_1I_3{\cal I}_1^2
+3T^3I_3I_1^2{\cal I}_1
-3m^2TI_3{\cal I}_1^2{\cal I}_2
+{\cal I}_3{\cal I}_1^3+T^4I_1^3I_3-6m^2T^2I_1I_3{\cal I}_1{\cal I}_2
\right.\\&&\left.
+3TI_1{\cal I}_1^2{\cal I}_3
-3m^2T^3I_1^2I_3{\cal I}_2
+3m^4TI_3{\cal I}_1^2{\cal I}_3
+3m^4TI_3{\cal I}_1{\cal I}_2^2
\right]\;,\\ \nonumber
{\cal F}_{\rm 3c}&=&{1\over24}g^6\left[
\left({\cal I}_1+TI_1-m^2{\cal I}_2\right)T^3I_{\rm ball}^{\prime}
+({\cal I}_1+TI_1){\cal I}_{\rm ball}^{\prime}
+3T^2I_1I_2{\cal I}_1{\cal I}_2
-m^2TI_2{\cal I}_2{\cal I}_{\rm sun}
\right. \\ &&\left.
+3T^3I_1^2I_2{\cal I}_2+2{\cal I}_1\left(
TI_1-m^2TI_2
\right)\sumint_{QR}{Q^2+(2/d){\bf q}^2\over Q^6R^2(Q+R)^2}
\right]\;,\\
{\cal F}_{\rm 3d}&=&{1\over48}g^6\left[
\sumint_P\left[\Pi(P)\right]^3
+T^4I_{\rm triangle}+6TI_1\sumint_P{1\over P^2}\left[\Pi(P)\right]^2
+3T^3{\cal I}_2I_{\rm ball}
\right]\;,\\ \nonumber
{\cal F}_{\rm 3e}&=&-{1\over8}g^4m_1^2\left[
T^2I_2^2{\cal I}_1+T^3I_1I_2^2
+2TI_2{\cal I}_1{\cal I}_2-2m^2TI_2{\cal I}_2^2
+{\cal I}_1{\cal I}_2^2
+2T^2I_1I_2{\cal I}_2-m^2T^2I_2^2{\cal I}_2
\right.\\ &&\left.
+TI_1{\cal I}_2^2
-4m^2TI_2{\cal I}_1{\cal I}_3
\right]\;,\\ \nonumber
{\cal F}_{\rm 3f}&=&-{1\over8}g^4m_1^2\left[
TI_3{\cal I}_1^2
+2T^2I_3I_1{\cal I}_1
+T^3I_3I_1^2
-2m^2TI_3{\cal I}_1{\cal I}_2+
{\cal I}_3{\cal I}_1^2
-2T^2m^2I_3I_1{\cal I}_2
+2TI_1{\cal I}_1{\cal I}_3
\right. \\ &&\left.
+2m^4TI_3{\cal I}_1{\cal I}_3
+m^4TI_3{\cal I }_2^2
\right]\;,\\ 
{\cal F}_{\rm 3g}&=&-{1\over12}g^4m_1^2\left[
T^3I_{\rm ball}^{\prime}
+{\cal I}_{\rm ball}^{\prime}
+
+3T^2I_1I_2{\cal I}_2
+2\left(TI_1-m^2TI_2\right)
\sumint_{QR}{Q^2+(2/d){\bf q}^2\over Q^6R^2(Q+R)^2}
\right]\;,
\\
{\cal F}_{\rm 3h}&=&{1\over4}g^2m_1^4\left[
TI_3{\cal I}_1+T^2I_1I_3
+TI_1{\cal I}_3
-m^2T{\cal I}_2I_3
+{\cal I}_1{\cal I}_3
+m^4T{\cal I}_3I_3
\right]\;, \\
{\cal F}_{\rm 3i}&=&{1\over8}g^2m_1^4
\left[T^2I_2^2+2TI_2{\cal I}_2+{\cal I}_2^2
-4m^2TI_2{\cal I}_3
\right]\;, \\
{\cal F}_{\rm 3j}&=&-{1\over6}m_1^6\left[TI_3+{\cal I}_3
\right] \;,
\label{last4}
\eqa
where the self-energy $\Pi(P)$ is defined in Eq.~(\ref{fullself})
and the integrals $I_{\rm ball}^{\prime}$ and $I_{\rm triangle}$ are
defined in Appendix B.
The poles in Eqs.~(\ref{first4})--(\ref{last4}) are cancelled by the
counterterms in Eq.~(\ref{fireloop}). The final result for the 
four-loop contribution to the free energy is 

\bqa\nonumber
    \mathcal F_3 &=&
        \frac{\pi^2 T^4}{90} \frac{5}{288} \frac{\alpha^3}{\hat m^3} \bigg[
            1
            + 18 \bigg(
                11 L + 3 \gamma_E - 6 - 16 \log 2 - 16 \log \hat m
                - 8 \frac{\zeta'(-1)}{\zeta(-1)}
                \bigg) \hat m^2
            \nonumber \\
            && \qquad
            + \bigg(
                1236 + 108C^a_{\rm triangle} + 36 C_{\rm ball}^{\prime}
 + 288 \gamma_1 - \frac{9198}{5} \gamma_E
                + 450 \gamma_E^2 - \frac{6456}{5} L + 432 \gamma_E L
                + 648 L^2 + 135 \pi^2
                \nonumber \\
                && \qquad
                - 54 \pi^2 C_{\rm triangle}^b- 216 \pi^2 \gamma_E
                + (2100 - 72 \gamma_E + 1728 L) \frac{\zeta'(-1)}{\zeta(-1)}
                + 432 \left(\frac{\zeta'(-1)}{\zeta(-1)}\right)^2
                - 432 (\gamma_E + 2 L) \frac{\zeta'(-3)}{\zeta(-3)}
                \nonumber \\
                && \qquad
                + 360\frac{\zeta''(-1)}{\zeta(-1)} + 1728 \log 2
                + 216 \pi^2 \log 2 + 432 (4 - \pi^2) \log \hat m - 4534 \zeta(3)
                \bigg) \hat m^3
            \nonumber \\
            && \qquad
            + \frac{9}{2} \bigg(
                3742 - 288 C_I - 48 C_{\rm ball}^{\prime}
 - 8064 \gamma_1 - 6072 \gamma_E
                - 2544 \gamma_E^2 - 3904 L - 1872 \gamma_E L
                \nonumber \\
                && \qquad
                - 2184 L^2 + 900 \pi^2
                + (1808 + 1824 \gamma_E + 2496 L) \frac{\zeta'(-1)}{\zeta(-1)}
                - 288 \frac{\zeta''(-1)}{\zeta(-1)} + 2688 \gamma_E \log 2
                \nonumber \\
                && \qquad
                + 4992 L \log 2 + 4992 (\gamma_E + L) \log \hat m
                - 2304 \gamma_E \log \pi + 2304 \log^2(2 \pi) - 15 \zeta(3)
                \bigg) \hat m^4
            \bigg]
            \nonumber \\
            &&
        -\frac{\pi^2 T^4}{90} \frac{5}{16} \frac{\alpha^2 \hat m_1^2}{\hat m^3} \bigg[
            1
            + \left(84 L + 36 \gamma_E - 96 \log \hat m - 36 - 96 \log 2
                    - 48 \frac{\zeta'(-1)}{\zeta(-1)}\right) \hat m^2
            \nonumber \\
            && \qquad
            + 2 \bigg(48(L+\gamma_E)^2 - 268(L+\gamma_E) 
- \gamma_E(17-21\gamma_E)
                    +  48 \gamma_1 -34 - \frac{9 \pi^2}{2}
                    \nonumber \\
                    && \qquad
                    - \frac{\zeta'(-1)}{\zeta(-1)}(34+12\gamma_E)
                    - 12 \frac{\zeta''(-1)}{\zeta(-1)}
                    + \zeta(3) + 6 C_{\rm ball}' \bigg) \hat m^3
            - 3 \bigg(89+120(L+\gamma_E)+[18(L+\gamma_E)]^2
                    \nonumber \\
                    && \qquad
+ 15 \zeta(3)\bigg) \hat m^4
            \bigg]
            \nonumber \\
            &&
        + \frac{\pi^2 T^4}{90} \frac{15}{8} \frac{\alpha \hat m_1^4}{\hat m^3} \bigg[
            1
            + 18 (L + \gamma_E) \hat m^2
            + [24 (L + \gamma_E)^2 + 2 \zeta(3)] \hat m^3
            - \frac{45}{2} \zeta(3) \hat m^4
            \bigg]
            \nonumber \\
            &&
        -\frac{\pi^2 T^4}{90} \frac{15}{4} \frac{\hat m_1^6}{\hat m^3} \bigg[
            1
            + 2\zeta(3) \hat m^3
            \bigg]\;,
 \label{4free}
\eqa
where the constants are 
\begin{eqnarray}
    C'_\mathrm{ball} &=& 48.7976\;, \\
    C_\mathrm{triangle}^a &=& - 25.7055\;, \\
    C_\mathrm{triangle}^b &=& 28.9250\;, \\
    C_{I} &=& - 38.5309\;.
\end{eqnarray}
There are a couple of calculational details that are worth while pointing out.
The $g^6$ contribution arising from diagram ${\cal F}_{\rm 3d}$
when all momenta are hard reads 
\bqa
{\cal F}_{\rm 3d}^{\rm (hhhh)}&=&\sumint_{P}[\Pi(P)]^3\;.
\eqa
This term can be combined with the $g^6$ term arising from 
the counterterm 
${\cal F}_{\rm 2b}\Delta_1g^2=-g^4{\cal I}_{\rm ball}\Delta_1g^2/48$
and gives
\bqa
\sumint_{P}\left\{[\Pi(P)]^3-{3\over(4\pi)^2\epsilon}[\Pi(P)]^2\right\}\;.
\eqa
This particular combination was first calculated 
by Gynther {\it et al.}~\cite{4loopfinns} using the
methods of Arnold and Zhai. Similarly, we combine the $g^7$ term
from ${\cal F}_{\rm 3d}$ with the term $TI_1{\cal I}_{\rm sun}$
from ${\cal F}_{\rm 2b}\Delta_1g^2$, which gives
\bqa
\sumint_{P}{1\over P^2}
\left\{[\Pi(P)]^2-{2\over(4\pi)^2\epsilon}[\Pi(P)]\right\}\;.
\eqa
We calculate this sum-integral in Appendix D.
Finally, the term from ${\cal F}_{\rm 2b}\Delta_1m^2$ which 
involves ${\cal I}_{\rm sun}$ can be combined with the term
$-m^2I_2{\cal I}_{2}{\cal I}_{\rm sun}$ arising from ${\cal F}_{3c}$ to give
\bqa
{1\over24}g^6m^2I_2\left({1\over(4\pi)^2}{1\over\epsilon}
-{\cal I}_2\right){\cal I}_{\rm sun}\;.
\label{combi2}
\eqa
Since ${\cal I}_{\rm  sun}$ vanishes at order $\epsilon^0$
and the term inside the paranthesis is finite, the particular
combination~(\ref{combi2}) vanishes in the limit $\epsilon\rightarrow0$.

\subsection{Pressure to four loops}
The pressure ${\cal P}$ is given by $-{\cal F}$. 
The contributions to the pressure of zeroth, first, second order, and
third order in
$g^2$ are given by Eqs.~(\ref{f0}),~(\ref{2free}),~(\ref{3free}), 
and~(\ref{4free}), 
respectively. Adding
these contributions and setting ${\cal E}_0=0$ and $m_1^2=m^2$, we obtain 
approximations to the pressure in screened perturbation theory which are 
accurate to ${\cal O}\left(g^7\right)$.

\noindent
The one-loop approximation to the pressure is
\bqa
{\cal P}_{0} &=&
{\cal P}_{\rm ideal} \Bigg[
1-15\hat{m}^2+60\hat{m}^3+45\hat{m}^4(L+\gamma_E)
-{15\over2}\zeta(3)\hat{m}^6
\Bigg]
\;,
\label{1p}
\eqa
where ${\cal P}_{\rm ideal}=\pi^2T^4/90$ is the pressure of an ideal gas 
of massless particles.

\noindent
The two-loop approximation to the pressure is obtained by adding 
Eq.~(\ref{2free}) with $m_1^2=m^2$:
\bqa\nonumber
{\cal P}_{0+1}&=&
{\cal P}_{\rm ideal} \Bigg\{
1-{5\over4}\alpha
+15\hat{m}\alpha
+15\hat{m}^2(L+\gamma_E-3)\alpha
-30\hat{m}^3\bigg[1+3(L+\gamma_E)\alpha\bigg]
\\  &&
-45\hat{m}^4\left[(L+\gamma_E)
+\left((L+\gamma_E)^2+{1\over12}\zeta(3)\right)\alpha\right]
+{45\over2}\zeta(3)\hat{m}^5\alpha
+15\zeta(3)\hat{m}^6
\Bigg\}
\;.
\label{2p}
\eqa
   
\noindent
The three-loop approximation to the pressure is obtained by adding
Eq.~(\ref{3free}) with $m_1^2=m^2$:
\bqa\nonumber
{\cal P}_{0+1+2}&=&
{\cal P}_{\rm ideal} \Bigg\{
1
+{5\over8\hat{m}}\alpha^2 
-{5\over4}\alpha
+\Bigg(
-{59\over12}+{15\over4}L+{5\over4}\gamma_E
+5{\zeta^{\prime}(-1)\over\zeta(-1)}
-{5\over2}{\zeta^{\prime}(-3)\over\zeta(-3)}
\Bigg)\alpha^2
\\\nonumber
&& 
+{15\over2}\hat{m}\Bigg[
1-\Bigg(
5+3\gamma_E+7L
-8\log\hat{m}
-8\log2
-4{\zeta^{\prime}(-1)\over\zeta(-1)}
\Bigg)\alpha
\Bigg]\alpha
+
       {5\over8} \hat m^2\bigg(
            268 (L + \gamma_E) - 48 (L + \gamma_E)^2
    \nonumber \\    && \nonumber
            + \frac{\zeta'(-1)}{\zeta(-1)} (34 + 12 \gamma_E)
            + 12 \frac{\zeta''(-1)}{\zeta(-1)}
            + \gamma_E (17 - 21 \gamma_E)
            + 34 + \frac{9 \pi^2}{2} - 48 \gamma_1
            - \zeta(3)
            - 6 C_{\rm ball}'
        \bigg)\alpha^2
\\ &&\nonumber
-{15\over2}\hat{m}^3\Bigg[1-6(L+\gamma_E)\alpha
-{1\over12}\bigg(89 + 120 (L+\gamma_E) + [18 (L+\gamma_E)]^2 + 15 \zeta(3) \bigg)\alpha^2 
\Bigg]
\\ &&
+45\hat{m}^4
\left((L+\gamma_E)^2+{1\over12}\zeta(3)\right)
\alpha
-{135\over4}\hat{m}^5\alpha\zeta(3)
-{15\over2}\hat{m}^6\zeta(3)
\Bigg\}
\label{3p}
\;.
\eqa
%

The four-loop approximation to the pressure is obtained by adding 
Eq.~(\ref{4free}) to Eq.~(\ref{3p}), with $m_1^2=m^2$:  
\begin{eqnarray}
    \frac{\mathcal P_{0+1+2+3}}{\mathcal P_\mathrm{ideal}} &=&
        1
        - \frac{5}{288} \frac{\alpha^3}{\hat m^3}
        + \frac{15}{16} \frac{1}{\hat m} \left[
            \alpha^2
            + \frac{1}{3} \left(
                16 \log \hat m + 6 - 3 \gamma_E - 11 L
                + 8 \frac{\zeta'(-1)}{\zeta(-1)} + 16 \log 2
                \right) \alpha^3
            \right]
        \nonumber \\
        &&
        - \frac{5}{4} \bigg[
            \alpha
            - \left(
                3 L - \frac{59}{15} \gamma_E
                + 4 \frac{\zeta'(-1)}{\zeta(-1)}
                - 2 \frac{\zeta'(-3)}{\zeta(-3)}
                \right) \alpha^2
            + \frac{1}{72} \bigg(
                1236 + 36 C'_\mathrm{ball} + 108 C_\mathrm{triangle}^a
                \nonumber \\
                && \qquad
                + 288 \gamma_1 - \frac{9198}{5} \gamma_E + 450 \gamma_E^2
                - \frac{6456}{5} L + 432 \gamma_E L
                + 648 L^2 + 135 \pi^2
                - 54 \pi^2 C_\mathrm{triangle}^b
                \nonumber \\
                && \qquad
                - 216 \pi^2 \gamma_E
                + (2100 - 72 \gamma_E + 1728 L) \frac{\zeta'(-1)}{\zeta(-1)}
                + 432 \left(\frac{\zeta'(-1)}{\zeta(-1)}\right)^2
                - 432 (\gamma_E + 2 L) \frac{\zeta'(-3)}{\zeta(-3)}
                \nonumber \\
                && \qquad
                + 360 \frac{\zeta''(-1)}{\zeta(-1)} + 1728 \log 2
                + 216 \pi^2 \log 2 + 432 (4 - \pi^2) \log \hat m - 4534 \zeta(3)
                \bigg) \alpha^3
            \bigg]
        \nonumber \\
        &&
        + \frac{45}{8} \hat m \bigg[
            \alpha
            - \frac{2}{3} \left(
                13 + 3 \gamma_E + 7 L - 4 \frac{\zeta'(-1)}{\zeta(-1)}
                - 8 \log 2 - 8 \log \hat m \right) \alpha^2
            - \frac{1}{72} \bigg(
                3742 - 288 C_I
                \nonumber \\
                && \qquad
                - 48 C'_\mathrm{ball}
                - 8064 \gamma_1 - 6072 \gamma_E
                - 2544 \gamma_E^2 - 3904 L - 1872 \gamma_E L
                - 2184 L^2 + 900 \pi^2
                \nonumber \\
                && \qquad
                + (1808 + 1824 \gamma_E + 2496 L) \frac{\zeta'(-1)}{\zeta(-1)}
                - 288 \frac{\zeta''(-1)}{\zeta(-1)} + 2688 \gamma_E \log 2
                + 4992 L \log 2
                \nonumber \\
                && \qquad
                + 4992 (\gamma_E + L) \log \hat m
                - 2304 \gamma_E \log \pi + 2304 \log^2(2 \pi) - 15 \zeta(3)
                \bigg) \alpha^3
            \bigg]
            \nonumber \\
            &&
        - \frac{15}{4} \hat m^3 \bigg[
            1
            - 3 (L + \gamma_E) \alpha
            + \frac{1}{12} \bigg(
                89 + 120 (L + \gamma_E) + [18 (L + \gamma_E)]^2 + 15 \zeta(3)
                \bigg) \alpha^2
            \bigg]
            \nonumber \\
            &&
        + \frac{135}{16} \zeta(3) \hat m^5 \alpha\;.
\label{resultfinale}
\end{eqnarray}

The final result for the pressure is given by Eq.~(\ref{resultfinale}).
If we use the weak-coupling expansion for the mass parameter, 
$\hat{m}^2=\alpha/6$, our result reduces to the weak-coupling expansion
result through order $\alpha^3$.~\footnote{It is important to point out
that we have only calculated part of the $g^7$-term in the weak-coupling 
expansion. See the discussion in Sec.~\ref{summary}.}
Inserting $\hat{m}^2$ into
Eq.~(\ref{resultfinale}), we obtain
\begin{eqnarray}
    \mathcal P &=& \mathcal P_\mathrm{ideal} \bigg\{
        1
        - \frac{5}{4} \alpha
        + \frac{5 \sqrt 6}{3} \alpha^{3/2}
        + \frac{15}{4} \bigg[
            \log\frac{\mu}{4 \pi T} + C_4
            \bigg] \alpha^2
        - \frac{15 \sqrt 6}{2} \bigg[
            \log\frac{\mu}{4 \pi T} - \frac{2}{3} \log \alpha + C_5
            \bigg] \alpha^{5/2}
    \nonumber \\
    &&
        - \frac{45}{4} \bigg[
            \log^2 \frac{\mu}{4 \pi T}
            - \frac{1}{3} \left(
                \frac{269}{45} - 2 \gamma_E
                - 8 \frac{\zeta'(-1)}{\zeta(-1)}
                + 4 \frac{\zeta'(-3)}{\zeta(-3)}
                \right) \log\frac{\mu}{4 \pi T}
            + \frac{1}{3} (4 - \pi^2) \log\alpha + C_6
            \bigg] \alpha^3
        \bigg\}\;,
\label{weak66}
\end{eqnarray}
where the constants $C_4$--$C_6$ are 
\begin{eqnarray}
    C_4 &\equiv&
        - \frac{59}{45} + \frac{1}{3} \gamma_E
        + \frac{4}{3} \frac{\zeta'(-1)}{\zeta(-1)}
        - \frac{2}{3} \frac{\zeta'(-3)}{\zeta(-3)}\;,
        \\
    C_5 &\equiv&
        \frac{5}{6}
        + \frac{1}{3} \gamma_E - \frac{2}{3} \log \frac{2}{3}
        - \frac{2}{3} \frac{\zeta'(-1)}{\zeta(-1)}\;,
        \\
    C_6 &\equiv&
        \frac{1}{3} (4 - \pi^2) \log\frac{2}{3}
        + \frac{103}{54} + \frac{1}{18} C'_\mathrm{ball} - \frac{1}{6} C_\mathrm{triangle}^a - \frac{\pi^2}{12} C_\mathrm{triangle}^b
        + \frac{4}{9} \gamma_1 - \frac{511}{180} \gamma_E + \frac{25}{36} \gamma_E^2
        + \frac{5 \pi^2}{24} - \frac{\pi^2}{3} \gamma_E
        \nonumber \\
        &&
        + \pi^2 \log 2
        + \left(\frac{175}{54} - \frac{1}{9} \gamma_E\right) \frac{\zeta'(-1)}{\zeta(-1)}
        + \frac{2}{3} \left(\frac{\zeta'(-1)}{\zeta(-1)}\right)^2
        + \frac{5}{9} \frac{\zeta''(-1)}{\zeta(-1)}
        - \frac{2}{3} \gamma_E \frac{\zeta'(-3)}{\zeta(-3)}
-{2267\over324}\zeta(3)\;.
\end{eqnarray}
The numerical values of $C_4$--$C_6$ are 
\begin{eqnarray}
    C_4 &=& 1.09775\;, \\
    C_5 &=& -0.0273205\;, \\
    C_6 &=& -6.59363\;.
\end{eqnarray}
Gynther {\it et al.}~\cite{4loopfinns} have calculated the pressure 
for an $O(N)$-symmetric theory at weak coupling through order $g^6$ 
using effective field theory methods.
Our result agrees with theirs for $N=1$.

Using the renormalization group equation for the running coupling constant
to next-to-leading order,
\bqa
\mu{d\alpha\over d\mu}&=&3\alpha^2-{17\over3}\alpha^3\;,
\eqa
it is straightforward to verify that the result~(\ref{weak66}) is
independent of the renormalization scale $\mu$ through order $g^6\log g$.

\section{Gap equations and Numerical results}
The mass parameter $m$ in screened perturbation theory is completely arbitrary.
In order to complete a calculation using SPT, we need a prescription for
the mass parameter $m$ as a function of $g$ and $T$.
One of the complications which arises from the ultraviolet divergences
is that the parameters ${\cal E}_0$, $m^2$, $m_1^2$, and $g^2$ are all
running parameters that depend on the renormalization scale $\mu$.

The prescription of Karsch, Patk\'os, and Petreczky 
for $m_*(T)$ is the solution to the
one-loop gap equation:
\bqa
\label{pet}
m_{*}^2={1\over2}\alpha(\mu_*)\left[
J_1(\beta m_*)T^2-\left(2\log{\mu_*\over m_*}+1\right)m_*^2
\right]
\;,
\eqa
%
where $\mu^*$ is the renormalization scale and $J_1(\beta m)$ is the function
\bqa
J_1(\beta m)&=&
8\beta^2\int_0^{\infty}{dpp^{2}\over(p^2+m^2)^{1/2}}
{1\over e^{\beta(p^2+m^2)^{1/2}}-1}\;.
\eqa
Their choice for the scale was $\mu_*=T$.
In the weak-coupling limit, the solution to~(\ref{pet}) is 
$m_*=g(\mu_*)T/\sqrt{24}$. 
The gap equation~(\ref{pet}) is the renormalized version of the following
equation
\bqa
m^2&=&{1\over2}g^2\sumint_P{1\over P^2+m^2}\;.
\label{ex2pin}
\eqa
There are many possibilities for generalizing~(\ref{pet}) to higher orders
in $g$. We will consider three different possibilities in the following.

\subsubsection{Debye mass}

One class of possibilities 
is to identify $m_*$
with some physical mass in the system.  The simplest choice is the 
Debye mass $m_D$ defined by the location of the pole in the static
propagator:
\bqa
p^2+m^2+\Sigma(0,p)&=&0\;,\hspace{1cm}p^2=-m_D^2\;.
\eqa
%
The Debye mass is a well defined quantity in scalar field theory and
abelian gauge theories at any order in perturbation theory. However,
in nonabelian gauge theories, it is plagued by infrared divergences beyond
leading order~\cite{reb}. 

\subsubsection{Tadpole mass}

The {\it tadpole mass} is another generalization of Eq.~(\ref{pet})
to higher loops.
It can be calculated by taking the partial derivative
of the free energy ${\cal F}$ with respect to $m^2$ {\it before}
setting $m_1=m$:
\bqa
m_t^2&=&g^2{\partial {\cal F}\over\partial m^2}\bigg|_{m_1=m}\;.
\eqa
From this equation, we see that $m_t^2$ is proportional to the
expectation value $\langle \phi^2\rangle$.
The tadpole mass is well defined at all orders in scalar field theory, but
the generalization to gauge theories is problematic. The natural replacement
of $\langle\phi^2\rangle$ would be $\langle A_{\mu}A_{\mu}\rangle$, which
is a gauge-variant quantity.

\subsubsection{Variational mass}

There is another class of prescriptions that is variational in spirit.
The results of SPT would be independent of $m$
if they were calculated to all orders. This suggests choosing $m$ to
minimize the dependence of some physical quantity on $m$.
The {\it variational mass} is defined by minimizing the free energy;
\bqa
{\partial {\mathcal F}\over \partial m^2}&=&0\;.
\eqa
%
The variational mass has the benefit that it is well defined at 
all orders in perturbation theory and can easily be generalized to 
gauge theories.

\subsubsection{Comparison}
At one loop, the three different prescriptions give the same 
gap equation, Eq.~(\ref{pet}). Moreover, it turns out that the two-loop
tadpole mass coincides with the one-loop tadpole mass~\cite{abssc}.  
However, at 
two loops the screening and variational masses are 
ill-behaved~\cite{abssc}.  The
screening mass solution ceases to exist beyond $g \sim 2.6$ and the
variational gap equation only has solutions in the vicinity of $g=0$
for some values of $L$.  
In the following, we therefore restrict ourselves to the tadpole gap equation.
\subsubsection{Tadpole gap equation through three loops}
At one loop, the renormalized gap equation follows from 
Eq.~(\ref{f0}) upon differentiation with respect to $m^2$ and 
can be written as
\bqa
0&=&\hat{m}^2
-{1\over6}\alpha
\left[1-6\hat{m}-6\hat{m}^2\left(L+\gamma_E\right)
+{3\over2}\zeta(3)\hat{m}^4
\right]\;.
\label{gap111}
\eqa
At two loops, the renormalized gap equation follows from differentiating
the sum of Eqs.~(\ref{f0}) and~(\ref{2free})
with respect to $m$, and setting $m_1=m$. It can be written in the form
\begin{eqnarray}
    0 &=&
    \hat{m}^2
    + {\alpha^2 \over 12 \hat m}
    - {\alpha \over 6} \bigg[1 + \alpha (3 - \gamma_E - L)\bigg]
    + {1 \over 2} \hat m \alpha \bigg[1 - 3 \alpha (\gamma_E + L)\bigg]
    \nonumber \\ &&
    - \hat m^2 \alpha^2 \bigg[(\gamma_E + L)^2 + {\zeta(3) \over 12}\bigg]
    + {5 \over 8} \hat m^3 \alpha^2 \zeta(3)
    + {1 \over 4} m^4 \alpha \zeta(3)\;.
    \label{2loopgap}
\end{eqnarray}
At three loops, the renormalized gap equation  
follows from differentiating 
the sum of Eqs.~(\ref{f0}),~(\ref{2free}), and~(\ref{3free}) and setting 
$m_1=m$. This yields
\bqa
    0 &=&
    \hat m^2
    + \frac{1}{8} \frac{\alpha^2}{\hat m} \bigg\{
        1
        + \alpha \bigg[
            1
            - \gamma_E
            - \frac{7}{3} L
            + \frac{4}{3} \frac{\zeta'(-1)}{\zeta(-1)}
            + \frac{8}{3} \log 2
            + \frac{8}{3} \log \hat m
            \bigg]
        \bigg\}
    \nonumber \\
    &&
    - \frac{\alpha}{6} \bigg\{
        1
        - \alpha (L + \gamma_E - 3)
        + \alpha^2 \bigg[
            2 (L + \gamma_E)^2
            - \frac{17}{12}
            + 2 \gamma_1
            - \frac{67}{6} (L + \gamma_E)
            - \frac{1}{24} \gamma_E (17 - 21 \gamma_E)
            \nonumber \\
            && \qquad \qquad
            - \frac{3 \pi^2}{16}
            - \frac{17}{12} \frac{\zeta'(-1)}{\zeta(-1)}
            - \frac{1}{2} \gamma_E \frac{\zeta'(-1)}{\zeta(-1)}
            - \frac{1}{2} \frac{\zeta''(-1)}{\zeta(-1)}
            + \frac{1}{24} \zeta(3)
            + \frac{1}{4} C'_\mathrm{ball}
            \bigg]
        \bigg\}
    \nonumber \\
    &&
    + \frac{3}{8} \hat m \alpha \bigg\{
        1
        - 2 \alpha (L + \gamma_E)
        + \alpha^2 \bigg[
            9 (L + \gamma_E)^2
            + {10 \over 3} (L + \gamma_E)
            + {89 \over 36}
            + {5 \over 12} \zeta(3)
            \bigg]
        \bigg\}
    - \frac{5}{16} \hat m^3 \alpha^2 \zeta(3)\;.
\label{gap333}
\eqa

\subsubsection{Numerical results}

\begin{figure}
    \center
\includegraphics[width=13.8cm]{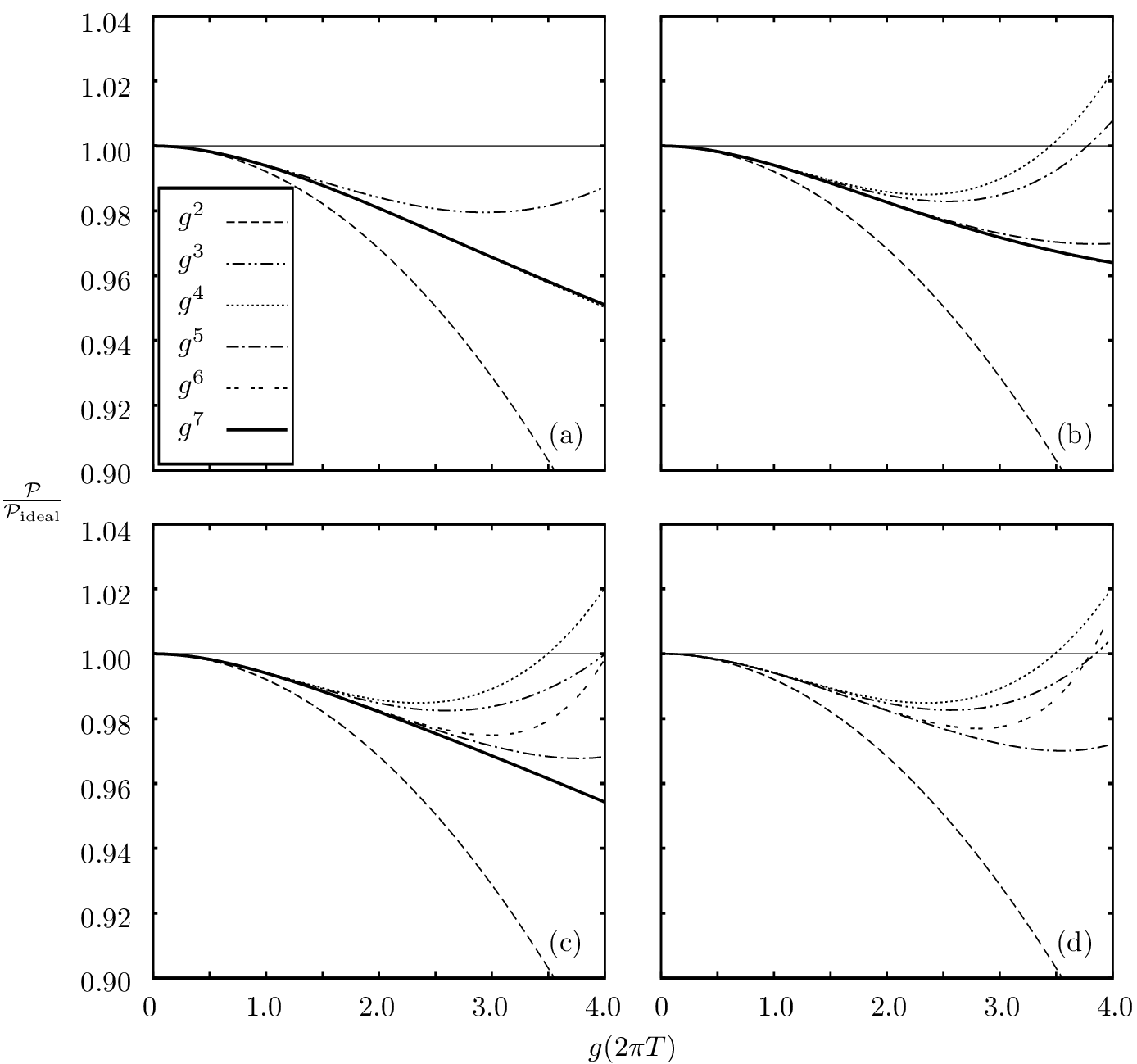}
    \caption{(a) 2-loop pressure, (b) 3-loop pressure, (c) 4-loop pressure, (d) weak-coupling expansion of the 
pressure, all normalized to ${\cal P}_{\rm ideal}$.}
\label{sptexp}
\end{figure}

The two-loop SPT-improved approximation to the
pressure is obtained by inserting the solution to the one-loop gap 
equation~(\ref{gap111}) into the two-loop pressure~(\ref{2p}).
In Fig.~\ref{sptexp}a we show the various truncations
to the two-loop SPT improved approximation to the 
${\cal P}/{\cal P}_{\rm ideal}$ 
as a function
of $g(2\pi T)$. We notice that the various truncations converge quickly.
The order-$g^4$ to
order-$g^7$ results are almost indistinguishable and essentially
equal to the exact numerical two-loop result in Ref.~\cite{abssc}.
In the three-loop case, we insert the solution to the two-loop gap
equation~(\ref{2loopgap}) into the three-loop pressure~(\ref{3p}).
In  Fig.~\ref{sptexp}b, we show the various truncations to the three-loop
SPT-improved approximation to 
${\cal P}/{\cal P}_{\rm ideal}$ as a function of $g(2\pi T)$ .
The three-loop result also converges to the exact numerical three-loop
result, albeit not as fast as in the two-loop case.
At four loops, we insert the solution to the three-loop gap
equation~(\ref{gap333}) into the four-loop pressure~(\ref{resultfinale}).
In  Fig.~\ref{sptexp}c, we show the various truncations to the four-loop
SPT-improved approximation to 
${\cal P}/{\cal P}_{\rm ideal}$ as a function of $g(2\pi T)$ .
Although we cannot compare our successive approximations with a numerically
exact 4-loop result for the pressure, we expect them to converge
reasonably fast. Based on the experience with the two- and three-loop
approximations, we expect that the $g^7$-truncation provides a good 
approximation to the numerically exact result. Clearly, however, only
a calculation through $g^8$ can settle this issue firmly.
In  Fig.~\ref{sptexp}d, we show the weak-coupling expansion of
${\cal P}/{\cal P}_{\rm ideal}$ to orders $g^2$, $g^3$, $g^4$, $g^5$, and $g^6$
as a function of $g(2\pi T)$ for comparison.
Note that the results to order $g^2$ are identical in SPT and in the
weak-coupling expansion since there is no $m$-dependence at this order.

\begin{figure}
    \center
\includegraphics[width=13.8cm]{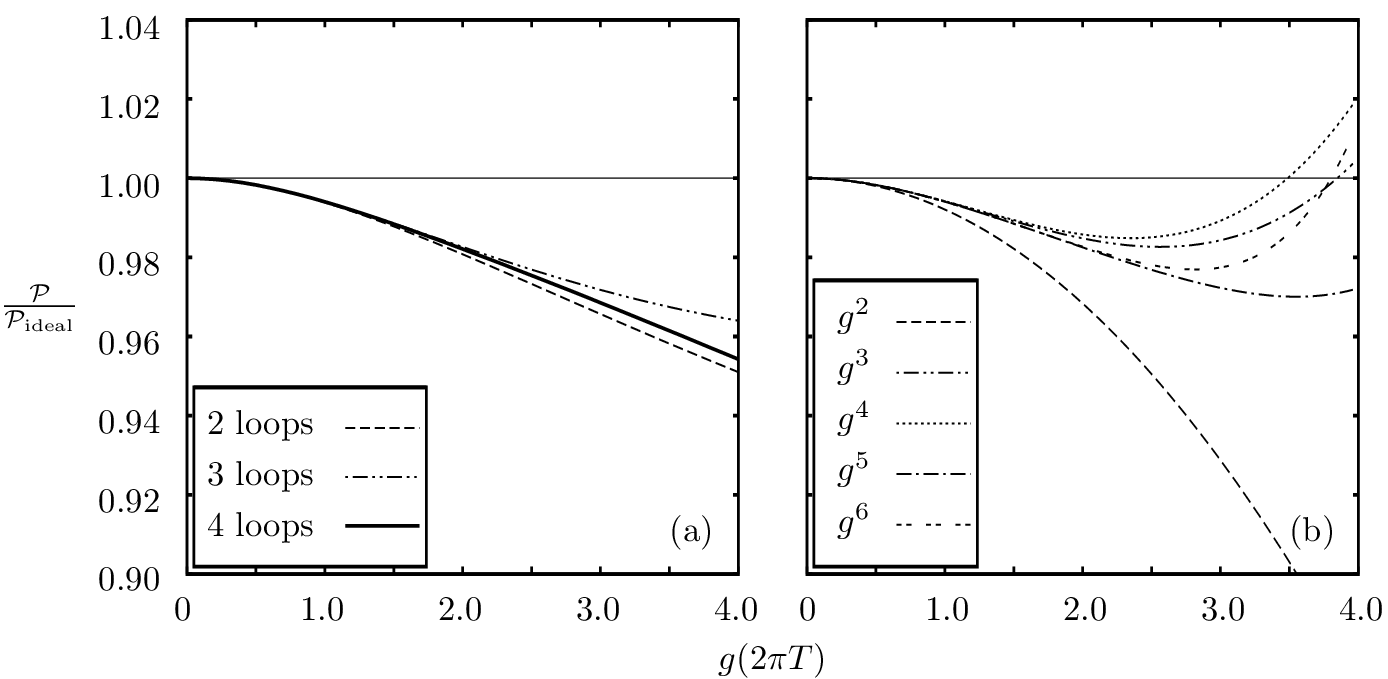}
    \caption{(a) Pressure normalized to ${\cal P}_{\rm ideal}$ 
through $g^7$ for various loop orders, (b) weak-coupling pressure at various 
orders of $g$.}
\label{sptweak}
\end{figure}

In Fig.\ref{sptweak}a, we show the two, three- and four-loop
pressure through order $g^7$ normalized to
${\cal P}/{\cal P}_{\rm ideal}$ as a function of $g(2\pi T)$ .
In  Fig.~\ref{sptweak}b, 
we show the weak-coupling expansion of
${\cal P}/{\cal P}_{\rm ideal}$ to orders $g^2$, $g^3$, $g^4$, $g^5$, and $g^6$
as a function of $g(2\pi T)$ for comparison.
The successive approximations using screened perturbation theory
have better convergence properties than the weak-coupling results.
The improved stability
is partly due to the fact that we are using a thermal mass determined by
a gap equation 
and not by the perturbative value for the Debye mass.

\section{Summary and Outlook}\label{summary}
In this paper, we have calculated the pressure of massless scalar field
theory to four loops using screened perturbation theory expanding
in a double expansion in powers of $g^2$ and $m/T$. Treating $m$
as ${\cal O}(gT)$, we truncated our expansion at order $g^7$.
The expansion required the evaluation of a new nontrivial three-loop
diagram, where we evaluated the sum-integral using the techniques developed
in Ref.~\cite{arnold1}.
We have seen that the successive approximations are more stable than
the weak-coupling expansion. In particular, it is interesting to 
note that the four-loop curve 
lies between the two-loop curve
and the three-loop curve. The apparent
improved convergence seemed to be linked to 
the fact that SPT basically is an expansion about an ideal gas of massive
particles instead of an expansion about an ideal gas of massless particles
which is the case for the weak-coupling expansion.

Using the weak-coupling value for the mass parameter $m$, our result 
reduces to the weak-coupling result for the pressure through $g^6$.
In particular, we have reproduced the pressure at weak coupling 
for $N=1$ obtained by Gynther {\it et al}~\cite{4loopfinns}.
Using effective-field theory methods , the authors
in Ref.~\cite{4loopfinns}
have calculated the hard and soft contributions to the pressure through
order $g^6$ separately. It appears that the convergence properties
in the hard sector are better than 
in the soft sector even for moderate values of the coupling.

We have mentioned that our result only includes part of the full
$g^7$-term in the weak coupling expansion. This is straightforward to
see if one uses the effective-field theory approach developed 
in~\cite{ericnieto1}. The contributions to the free energy comes from the two
momentum scales $T$ and $gT$. The contribution from the hard scale $T$
can be calculated by evaluating the sum-integrals with bare propagators and so
is therefore a series in $g^2$ starting at order $g^0$. 
The contribution to the free energy from the
soft scale $gT$ can be calculated using an effective 
Euclidean three-dimensional field
theory whose coefficients depend on $g$ and $T$. This contribution to the
free energy is a series in $g$ starting at $g^3$. The contributions to the
free energy that are odd in powers in $g$ is therefore entirely coming from 
three-dimensional vacuum diagrams
and power-counting tells you immediately 
that part of the $g^7$ term is arising from the
five-loop vacuum diagrams. Our four-loop calculation therefore agrees with
the weak-coupling expansion through order $g^6$.

In order to evaluate the free energy to order $g^7$, we must 
determine all the coefficients in the effective theory to sufficiently high
order in $g$. The only nontrivial calculation that is required is to determine
the mass parameter in the effective theory to order $g^6$. This involves
the expression for the diagram calculated in Appendix D i.e, 
the sum-integral
\bqa
I&\equiv&\sumint_P{1\over P^2}\left\{[{\Pi}(P)]^2-{2\over(4\pi)^2\epsilon}
{\Pi}(P)\right\}\;,
\eqa
The evaluation of the free energy to order $g^7$ is in progress~\cite{usnew}.

\section*{Acknowledgments}
J. O. A. would like to thank E. Braaten, M. Laine, and M. Strickland for
valuable discussions, and
thank E. Braaten and M. Strickland
for collaboration on related work on SPT.  

\appendix
\renewcommand{\theequation}{\thesection.\arabic{equation}}
\section{Sum-integrals}\label{appa}
In the imaginary-time formalism for thermal field theory, 
the 4-momentum $P=(P_0,{\bf p})$ is Euclidean with $P^2=P_0^2+{\bf p}^2$. 
The Euclidean energy $p_0$ has discrete values:
$P_0=2n\pi T$ for bosons,
where $n$ is an integer. 
Loop diagrams involve sums over $P_0$ and integrals over ${\bf p}$. 
With dimensional regularization, the integral is generalized
to $d = 3-2 \epsilon$ spatial dimensions.
We define the dimensionally regularized sum-integral by
\bqa
  \hbox{$\sum$}\!\!\!\!\!\!\int_{P}& \;\equiv\; &
  \left(\frac{e^\gamma\mu^2}{4\pi}\right)^\epsilon\;
  T\!\!\!\!\!\!\sum_{P_0=2n\pi T}\:\int {d^{3-2\epsilon}p \over (2 \pi)^{3-2\epsilon}}\;,
\label{sumint-def}
\eqa
where $3-2\epsilon$ is the dimension of space and $\mu$ is an arbitrary
momentum scale. 
The factor $(e^\gamma/4\pi)^\epsilon$
is introduced so that, after minimal subtraction 
of the poles in $\epsilon$
due to ultraviolet divergences, $\mu$ coincides 
with the renormalization
scale of the $\overline{\rm MS}$ renormalization scheme.

\subsection{One-loop sum-integrals}
The massless one-loop sum-integral is given by 
\bqa\nonumber
{\cal I}_n&\equiv&
\sumint_{P}{1\over P^{2n}}\\
&=&(e^{\gamma_E}\mu^2)^{\epsilon}{\zeta(2n-3+2\epsilon)\over8\pi^2}
{\Gamma(n-\mbox{$3\over2$}+\epsilon)
\over\Gamma(\mbox{$1\over2$})\Gamma(n)} 
(2\pi T)^{4-2n-2\epsilon}\;,
\eqa
where $\zeta(x)$ is Riemann's zeta function.
Specifically, we need the sum-integrals
\bqa\nonumber
{\cal I}_0^{\prime}&\equiv&
\sumint_P\log P^2
\\ &=&
-{\pi^2T^4\over45}\left[1+{\cal O}\left(\epsilon\right)\right]\;,\\
{\cal I}_1
&=&{T^2\over12}\left({\mu\over4\pi T}\right)^{2\epsilon}
\left[1+\left(2+2{\zeta^{\prime}(-1)\over\zeta(-1)}\right)\epsilon
+\left(4+{\pi^2\over4}
+4{\zeta^{\prime}(-1)\over\zeta(-1)}
+2{\zeta^{\prime\prime}(-1)\over\zeta(-1)}
\right)\epsilon^2+
{\cal O}\left(\epsilon^3\right)\right]\;,\\
{\cal I}_2&=&
{1\over(4\pi)^2}\left({\mu\over4\pi T}\right)^{2\epsilon}
\left[{1\over\epsilon}
+2\gamma_E+\left({\pi^2\over4}-4\gamma_1\right)\epsilon\
+{\cal O}\left(\epsilon^2\right)\right]\;,\\
{\cal I}_3
&=&{1\over(4\pi)^4T^2}\left[2\zeta(3)+
{\cal O}\left(\epsilon\right)\right]\;.
\eqa

\subsection{Two-loop sum-integrals}
We need two two-loop sum-integral that are listed below:
\bqa\nonumber
{\cal I}_{\rm sun}&=&
\sumint_{PQ}{1\over P^2Q^2(P+Q)^2} \\
&=&{\cal O}(\epsilon)\;,\\
\sumint_{PQ}{P^2+(2/d)p^2\over P^6Q^2(P+Q)^2}&=&
{3\over4(4\pi)^4}\left({\mu\over4\pi T}\right)^{4\epsilon}
\left[{1\over\epsilon^2}+\left({5\over6}+4\gamma_E\right)
{1\over\epsilon}+{89\over36}+{\pi\over2}
+{10\over3}\gamma_E+4\gamma_E^2-8\gamma_1+{\cal O}(\epsilon)
\right]\;.
\label{sumintbrapet}
\eqa
The setting-sun sum-integral was first calculated by Arnold and Zhai in
Ref.~\cite{arnold1}, while Eq.\ \eqref{sumintbrapet} was calculated in
Ref.\ \cite{BP-01}.

\subsection{Three-loop sum-integrals}
We need the following three-loop sum-integrals:
\begin{widetext}
\bqa\nonumber
{\cal I}_{\rm ball}
&=&\sumint_{PQR}{1\over P^2Q^2R^2(P+Q+R)^2}\\
&=&{T^4\over24(4\pi)^2}\left({\mu\over4\pi T}\right)^{6\epsilon}\left[
{1\over\epsilon}+{91\over15}+8{\zeta^{\prime}(-1)\over\zeta(-1)}
-2{\zeta^{\prime}(-3)\over\zeta(-3)}
+{\cal O}(\epsilon)
\right]\;, 
\\ \nonumber
{\cal I}_{\rm ball}^{\prime}
&=&
\sumint_{PQR}{1\over P^4Q^2R^2(P+Q+R)^2} \\ \nonumber
&=&{T^2\over8(4\pi)^4}\left({\mu\over4\pi T}\right)^{6\epsilon}
\left[
{1\over\epsilon^2}+
\left({17\over6}+4\gamma_E
+2{\zeta^{\prime}(-1)\over\zeta(-1)}\right)
{1\over\epsilon}
\right.\\&&\left.
+{1\over2}
\gamma_E\left(17+15\gamma_E+12{\zeta^{\prime}(-1)\over\zeta(-1)}\right)
+C_{\rm ball}^{\prime}
+{\cal O}(\epsilon)
\right]\;, \\
\label{bp222}
\nonumber 
\sumint_P{1\over P^2}\left\{[\Pi(P)]^2-{2\over(4\pi)^2\epsilon}
\Pi(P)\right\}
    & = &
        -\frac{T^2}{4(4\pi)^4}\left({\mu\over4\pi T}\right)^{6\epsilon}
 \bigg\{
            \frac{1}{\epsilon^2}
            + \frac{1}{\epsilon} \left[
\frac{4}{3}+2\frac{\zeta'(-1)}{\zeta(-1)} + 4\gamma_E 
                \right]
    \nonumber \\    && \nonumber
            + \frac{1}{3} \bigg[46 - 8\gamma_E 
- 16\gamma_E^2 
- 104\gamma_1 - 24\gamma_E \log(2\pi)
                + 24\log^2(2\pi) + 
\frac{45\pi^2}{4} 
\\ &&
+ 24\frac{\zeta'(-1)}{\zeta(-1)}
+ 2\frac{\zeta''(-1)}{\zeta(-1)}
                + 16\gamma_E \frac{\zeta'(-1)}{\zeta(-1)}
                \bigg]
+C_I
+{\cal O}(\epsilon)            \bigg\}\;,
\label{oss}
\eqa
\end{widetext}
where $C_{\rm ball}^{\prime}=48.7976$ and $C_I=-38.5309$.
The massless basketball sum-integral was first calculated 
in Ref.~\cite{arnold1}, and ${\cal I}_{\rm ball}'$ in Ref.~\cite{4loopfinns}. 
The expression for the 
sum-integral Eq.~(\ref{oss}) appears here for the first time and is 
calculated in in appendix D.

\subsection{Four-loop sum-integrals}
We also need a single four-loop sum-integral which was calculated in
Ref.~\cite{4loopfinns}:
\begin{widetext}
\bqa\nonumber
\sumint_{P}\left\{[\Pi(P)]^3-{3\over(4\pi)^2\epsilon}[\Pi(P)]^2\right\}
&=&
-{T^4\over16(4\pi)^4}
\left[{1\over\epsilon^2}+\left(
{10\over3}+4{\zeta^{\prime}(-1)\over\zeta(-1)}+4L
\right){1\over\epsilon}
+(2L+\gamma_E)^2
\right.\\ &&\left.
\hspace{-4.8cm}
+\left({6\over5}-2\gamma_E
+4{\zeta^{\prime}(-3)\over\zeta(-3)}\right)(2L+\gamma_E)
+C_{\rm triangle}^a
\right]
-{T^4\over512(4\pi)^2}\left[
{1\over\epsilon}+8L
+4\gamma_E
+C_{\rm triangle}^b
\right]
+{\cal O}(\epsilon)\;,
\eqa
\end{widetext}
where $C_{\rm triangle}^a=-25.7055$ and $C_{\rm triangle}^b=28.9250$.

\section{Three-dimensional integrals}\label{appb}
Dimensional regularization can be used to
regularize both the ultraviolet divergences and infrared divergences
in 3-dimensional integrals over momenta.
The spatial dimension is generalized to  $d = 3-2\epsilon$ dimensions.
Integrals are evaluated at a value of $d$ for which they converge and then
analytically continued to $d=3$.
We use the integration measure
\begin{equation}
 \int_p\;\equiv\;
  \left(\frac{e^\gamma\mu^2}{4\pi}\right)^\epsilon\;
\:\int {d^{3-2\epsilon}p \over (2 \pi)^{3-2\epsilon}}\;.
\label{int-def}
\end{equation}

\subsection{One-loop integrals}
The one-loop integral is given by
\bqa\nonumber
I_n&\equiv&\int_p{1\over(p^2+m^2)^n}\\
&=&{1\over8\pi}(e^{\gamma_E}\mu^2)^{\epsilon}
{\Gamma(n-\mbox{$3\over2$}+\epsilon)
\over\Gamma(\mbox{$1\over2$})\Gamma(n)}m^{3-2n-2\epsilon}\;.
\eqa
Specifically, we need
\bqa\nonumber
I_0^{\prime}&\equiv&
\int_p\log(p^2+m^2)\\
&=&
-{m^3\over6\pi}\left({\mu\over2m}\right)^{2\epsilon}
\left[
1
+{8\over3}
\epsilon
+\left(
{52\over9}+{\pi^2\over4}\right)\epsilon^2
+{\cal O}
\left(\epsilon^3\right)
\right]\;,\\ 
I_1&=&-{m\over4\pi}\left({\mu\over2m}\right)^{2\epsilon}
\left[
1+2
\epsilon+\left(
4+{\pi^2\over4}\right)\epsilon^2
+{\cal O}\left(\epsilon^3\right)
\right]\;,\\
I_2&=&{1\over8\pi m}\left({\mu\over2m}\right)^{2\epsilon}
\left[
1
+{\pi^2\over4}
\epsilon^2
+{\cal O}\left(\epsilon^3\right)
\right]\;,\\
I_3&=&{1\over32\pi m^3}\left({\mu\over2m}\right)^{2\epsilon}
\left[
1+2
\epsilon
+{\pi^2\over4}\epsilon^2
+{\cal O}\left(\epsilon^3\right)
\right]\;.
\eqa
\subsection{Three-loop integrals}
We need two three-loop integrals:
\bqa\nonumber
I_{\rm ball}&=&
\int_{pqr}{1\over p^2+m^2}{1\over q^2+m^2}{1\over r^2+m^2}
{1\over({\bf p}+{\bf q}+{\bf r})^2+m^2}
\\ &=&
-{m\over(4\pi)^3}\left({\mu\over2m}\right)^{6\epsilon}
\left[{1\over\epsilon}+8-4\log2
+4\left(13+{17\over48}\pi^2-8\log2+\log^22\right)\epsilon
+{\cal O}\left(\epsilon^2\right)
\right]\;,
\\ \nonumber
I_{\rm ball}^{\prime}&=&
\int_{pqr}{1\over(p^2+m^2)^2}{1\over q^2+m^2}{1\over r^2+m^2}
{1\over({\bf p}+{\bf q}+{\bf r})^2+m^2} \\
&=&
{1\over8m(4\pi)^3}\left({\mu\over2m}\right)^{6\epsilon}
\left[{1\over\epsilon}+2-4\log2
+4\left(1+{17\over48}\pi^2-2\log2+\log^22\right)\epsilon
+{\cal O}\left(\epsilon^2\right)\right]\;.
\eqa
The massive basketball was calculated
in Ref.~\cite{ericnieto1} to order $\epsilon^0$, and to order
$\epsilon$ in Ref.~\cite{sunfinn}. The other 3-loop integral is obtained
by differentianting the massive basketball with respect to the mass $m$.
\subsection{Four-loop integrals}
We need a single four-loop integral, namely the triangle integral.
This integral was calculated in Ref.~\cite{sunfinn} and reads
\bqa\nonumber
I_{\rm triangle}&=&
\int_{pqrs}{1\over q^2+m^2}{1\over({\bf p}+{\bf q})^2+m^2}
{1\over r^2+m^2}{1\over({\bf p}+{\bf r})^2+m^2}
{1\over s^2+m^2}{1\over({\bf p}+{\bf s})^2+m^2}
\\
&=&{\pi^2\over32(4\pi)^4}\left({\mu\over2m}\right)^{8\epsilon}\left[
{1\over\epsilon}+2+4\log2
-{84\over\pi^2}\zeta(3)+{\cal O}\left(\epsilon\right)
\right]\;.
\eqa

\section{$m/T$ expansions}\label{appc}
In this appendix, we list the $m/T$ expansions of the sum-integrals we need.
The sum-integrals include sums over the Matsubara frequencies $P_0=2\pi n T$
and integrals over the three-momentum ${\bf p}$. In the sum-integrals, two
important mass scales appear. These are the {\it hard} scale $2\pi T$
and the {\it soft} scale $m$. The soft scale $m$ is of order $gT$
and at weak coupling this scale is well separated from the hard scale,
$m\ll 2\pi T$. We can therefore expand the sum-integrals as a 
Taylor series in powers of $m/T$.

First consider the simple one-loop sum-integral appearing in the 
expression for the one-loop free energy in Eq.~(\ref{f000}):
\bqa\nonumber
{\cal F}_{\rm 0a}&=&{1\over2}
\sumint_P\log\left[P^2+m^2\right]\\
&=&{1\over2}\sumint_P^{\rm (h)}\log\left[P^2+m^2\right]
+{1\over2}\sumint_P^{\rm (s)}\log\left[P^2+m^2\right]\;,
\eqa
where the superscripts $\rm (h)$ and $\rm (s)$ 
denote the hard and soft contributions,
respectively. In the hard region, the momentum $P$ is of order $T$ and so
we can expand in powers of $m^2/P^2$. This yields
\bqa
\sumint_P^{\rm (h)}\log\left[P^2+m^2\right]&=&
\sumint_P\log P^2+m^2\sumint_P{1\over P^2}-{1\over 2}m^4\sumint_P{1\over P^4}
+\cdots
\eqa
The contribution from soft momenta is given by the $p_0=0$ mode alone and
and reads
\bqa
\sumint_P^{\rm (s)}\log\left[P^2+m^2\right]
&=&T\int_p\log(p^2+m^2)\,.
\eqa
The other simple one-loop sum-integrals are expanded in a similar manner.

We next consider the massive basketball diagram in Eq.~(\ref{massen}):
\bqa
{\cal I}_{\rm ball}(m^2)
&=&\sumint_{PQR}{1\over(P^2+m^2)(Q^2+m^2)(R^2+m^2)[(P+Q+R)^2+m^2]}\;.
\label{masbask}
\eqa
Eq.~(\ref{masbask}) involves three sum-integrals
and so receives contributions from four momentum regions: 
$\rm (hhh)$, $\rm (hhs)$,
$\rm (hss)$, and $\rm (sss)$.
In the first case, where all the loop momenta are hard, 
we can expand the sum-integral in powers of
$m^2$. This yields
\bqa
{\cal I}_{\rm ball}^{\rm (hhh)}(m^2)
&=&\sumint_{PQR}{1\over P^2Q^2R^2(P+Q+R)^2}
-4m^2\sumint_{PQR}{1\over P^4Q^2R^2(P+Q+R)^2}+\cdots
\eqa
When two momenta are hard and one is soft, the contribution reads
\bqa\nonumber
{\cal I}_{\rm ball}^{\rm (hhs)}(m^2)
&=&4T\int_p{1\over p^2+m^2}
\sumint_{QR}{1\over Q^2+m^2}{1\over R^2+m^2}{1\over({\bf p}+Q+R)^2+m^2}
\\
&=&4T\int_p{1\over p^2+m^2}\sumint_{QR}{1\over Q^2R^2(Q+R)^2}
-8m^2T\int_p{1\over p^2+m^2}
\left[\sumint_{QR}{Q^2+(2/d){\bf q}^2\over Q^6R^2(Q+R)^2}\right]+\cdots\;.
\eqa
When one momentum is hard and two are soft, the contribution is
given by
\bqa\nonumber
{\cal I}_{\rm ball}^{\rm (hss)}(m^2)
&=&6T^2\int_{pq}{1\over p^2+m^2}{1\over q^2+m^2}
\sumint_R
{1\over R^2+m^2}{1\over{({\bf p}+{\bf q}+R)^2+m^2}} \\
&=&
6T^2\int_{pq}{1\over p^2+m^2}{1\over q^2+m^2}
\sumint_R{1\over R^4}+\cdots\;.
\eqa
Finally, when all momenta are soft, the contribution is given by the
massive basketball diagram $I_{\rm ball}$ in three dimensions:
\bqa
{\cal I}_{\rm ball}^{\rm (sss)}(m^2)&=&T^3
\int_{pqr}{1\over p^2+m^2}{1\over q^2+m^2}{1\over r^2+m^2}
{1\over({\bf p}+{\bf q}+{\bf r})^2+m^2}\;.
\eqa
The basketball diagram with a single mass insertion, 
${\cal I}_{\rm ball}^{\prime}(m^2)$, can be calculated
by differentiating the massive basketball diagram with respect to $m^2$.
This yields
\bqa\nonumber
{\cal I}_{\rm ball}^{\prime}(m^2)&=&
\sumint_{PQR}{1\over(P^2+m^2)^2}{1\over Q^2+m^2}{1\over R^2+m^2}
{1\over(P+Q+R)^2+m^2} \\ \nonumber
&=&
\sumint_{PQR}{1\over P^4Q^2R^2(P+Q+R)^2}
+T\int_p{1\over(p^2+m^2)^2}\sumint_{QR}{1\over Q^2R^2(Q+R)^2}
\\ &&
\nonumber
+2T\int_p{p^2\over(p^2+m^2)^2}
\left[\sumint_{QR}{Q^2+(2/d){\bf q}^2\over Q^6R^2(Q+R)^2}\right]
+3T^2\int_{pq}{1\over p^2+m^2}{1\over (q^2+m^2)^2}\sumint_R{1\over R^4}
\\ &&
+T^3
\int_{pqr}{1\over(p^2+m^2)^2}
{1\over q^2+m^2}{1\over r^2+m^2}{1\over({\bf p}+{\bf q}+{\bf r})^2+m^2}+\cdots\;.
\eqa 
Note that the second term is formally of order $g^5$, but it vanishes
at order $\epsilon^0$ due to the fact that 
${\cal I}_{\rm sun}={\cal O}(\epsilon)$.

The massive four-loop triangle sum-integral reads
\bqa
{\cal I}_{\rm triangle}(m^2)&=&
\sumint_{PQRS}{1\over Q^2+m^2}{1\over(P+Q)^2+m^2}
{1\over R^2+m^2}{1\over(P+R)^2+m^2}
{1\over S^2+m^2}{1\over(P+S)^2+m^2}\;.
\eqa
When all four momenta are hard, the leading contribution is given
by setting $m=0$, i. e.
\bqa
{\cal I}_{\rm triangle}^{\rm (hhhh)}(m^2)
&=&\sumint_{PQRS}{1\over Q^2(P+Q)^2R^2(P+R)^2S^2(P+S)^2}\;.
\eqa
When one of the momenta is hard and three are soft, we find
\bqa\nonumber
{\cal I}_{\rm triangle}^{\rm (hsss)}(m^2)
&=&3T^3\int_{pqr}
{1\over p^2+m^2}{1\over q^2+m^2}{1\over r^2+m^2}
{1\over({\bf p}+{\bf q}+{\bf r})^2+m^2}
\sumint_S{1\over S^4}+\cdots\;.
\eqa
This contribution is of order $g^7$.
When one momentum is soft and three momenta are hard, the contribution is
\bqa\nonumber
{\cal I}_{\rm triangle}^{\rm (shhh)}(m^2)
&=&6T\int_s{1\over s^2+m^2}\sumint_{PQR}
{1\over P^2+m^2}
{1\over Q^2+m^2}{1\over(P+Q)^2+m^2}
{1\over R^2+m^2}{1\over(P+R)^2+m^2} \\
&=&
6T\int_s{1\over s^2+m^2}\sumint_{PQR}
{1\over P^2Q^2R^2(P+Q)(P+R)^2}+\cdots\;.
\label{newdia}
\eqa
This contribution is of order $g^7$.
When all four loop momenta are soft, the contribution is given by the
massive three-dimensional triangle diagram $I_{\rm triangle}$:
\bqa
{\cal I}_{\rm triangle}^{\rm (ssss)}(m^2)&=&
T^4\int_{pqrs}
{1\over q^2+m^2}{1\over({\bf p}+{\bf q})^2+m^2}
{1\over r^2+m^2}{1\over({\bf p}+{\bf r})^2+m^2}
{1\over s^2+m^2}{1\over({\bf p}+{\bf s})^2+m^2}\;.
\eqa
This contribution is of order $g^6$.
Finally, we notice that the contribution when
two momenta are soft and two momenta
are hard, is of higher order in the coupling $g$.

\section{Explicit calculations}\label{appd}
In this appendix, we illustrate the use of the calculational techniques
developed by Arnold and Zhai in Ref.~\cite{arnold1} 
to evaluate complicated multiloop diagrams. The strategy is to rewrite
the original sum-integral into two sets of terms. The first type is
ultraviolet divergent, but is sufficiently simple to be evaluated analytically
using dimensional regularization. The second type is finite both in the
ultraviolet and the infrared, but is normally so complicated that it must be
evaluated numerically. In order to isolate the divergences in terms that
are tractable, typically one or more subtractions are required.

We need to calculate the following three-loop diagram
\bqa
\label{sumintdef}
I&\equiv&\sumint_P{1\over P^2}\left\{[{\Pi}(P)]^2-{2\over(4\pi)^2\epsilon}
{\Pi}(P)\right\}\;,
\eqa
where the self-energy ${\Pi}(P)$ is defined by
\bqa
\label{fullself}
{\Pi}(P)&=&\sumint_Q{1\over Q^2(P+Q)^2}\;.
\eqa
The first term in Eq.~(\ref{sumintdef}) arises from the $m/T$-expansion
of the triangle sum-integral in four dimensions, while the second term
arises from the term $TI_1{\cal I}_{\rm sun}$
which is a part of the counterterm ${\cal F}_{\rm 2b}\Delta_1g^2/g^2$ .

At zero temperature, the self-energy is denoted by ${\Pi}^0(P)$ 
and reads
\bqa
\label{pitildef}
{\Pi}^0(P)&=&
{1\over(4\pi)^2}\left({e^{\gamma_E}\mu^2\over P^2}\right)^{\epsilon}
{\Gamma(\epsilon)\Gamma^2(1-\epsilon)\over\Gamma(2-2\epsilon)}\;.
\eqa
In order to isolate the UV divergences and simplify the calculations, 
we write the self-energy as
\bqa
\Pi(P)&=&{1\over(4\pi)^2\epsilon}+{\Pi}^0_s(P)
+{\Pi}^T(P)\;,
\label{deco}
\eqa
where  ${\Pi}^0_s(P)$ is the finite part of ${\Pi}^0(P)$,
i.e. we have subtracted the divergent piece in Eq.~(\ref{pitildef})
from ${\Pi}^0(P)$:
\bqa
{\Pi}^0_s(P)&=&
{1\over(4\pi)^2}\left\{\left({e^{\gamma_E}\mu^2\over P^2}\right)^{\epsilon}
{\Gamma(\epsilon)\Gamma^2(1-\epsilon)\over\Gamma(2-2\epsilon)}
-{1\over\epsilon}\right\}\;,
\eqa
and ${\Pi}^T(P)$ is the finite-temperature piece of ${\Pi}(P)$. 
In three dimension, ${\Pi}^T(P)$ reads~\cite{arnold1}
\bqa
{\Pi}^T(P)&=&{T\over(4\pi)^2}\int{d^3r\over r^2}e^{i{\bf p}\cdot{\bf r}}
\left(\coth{\bar{r}}-{1\over\bar{r}}\right)
e^{-|p_0|r}\;,
\label{3dpt}
\eqa
where $\bar{r}=2\pi Tr$. In the following we need the UV limit of 
$\Pi^T(P)$. This happens to be given by the UV limit of the full 
self-energy~(\ref{fullself}) and is given by~\cite{arnold1}
\bqa
\Pi^T_{\rm UV}(P)&=&{2\over P^2}\sumint_Q{1\over Q^2}\;.
\eqa

Using the decomposition~(\ref{deco}), the integral in 
Eq.~(\ref{sumintdef}) can be written as
\bqa
I&=-&{1\over(4\pi)^4\epsilon^2}\sumint_P{1\over P^2}
+\sumint_P{1\over P^2}[{\Pi}_s^0(P)]^2
+2\sumint_P{1\over P^2}{\Pi}_s^0(P){\Pi}^T(P)
+\sumint_P{1\over P^2}[{\Pi}^T(P)]^2\;.
\label{deco2}
\eqa
We now consider the different contributions to $I$. The first term 
in Eq.~(\ref{deco2}) is a simple
one-loop sum-integral and reads:
\bqa\nonumber
I_1&=&-
{1\over(4\pi)^4\epsilon^2}\sumint_P{1\over P^2} \\ 
&=&-\left({\mu\over4\pi T}\right)^{2\epsilon}
{T^2\over12(4\pi)^4}\bigg[
{1\over\epsilon^2}+
2\left(
1+{\zeta^{\prime}(-1)\over\zeta(-1)}
\right){1\over\epsilon}
+{\pi^2\over4}
+4+4{\zeta^{\prime}(-1)\over\zeta(-1)}
+2{\zeta^{\prime\prime}(-1)\over\zeta(-1)}
+{\cal O}(\epsilon)
\bigg]\;.
\label{adding2}
\eqa
The second term in Eq.~(\ref{deco2}) 
contains no logarithmic UV divergences and so it is
finite in dimensional regularization:
\bqa\nonumber
I_2
&=&
\sumint_P{1\over P^2}[{\Pi}_s^0(P)]^2 \\ 
&=&
    {T^2 \over 12 (4\pi)^4}
    \bigg[ 4 + {\pi^2 \over 3}
+
    8{\zeta^{\prime}(-1) \over \zeta(-1)}
    \left(2 + \log{\mu \over 4\pi T}\right)
+
    4{\zeta^{\prime\prime}(-1) \over \zeta(-1)}
    + 4 \left( 2 + \log{\mu \over 4\pi T} \right)^2
\bigg]
+\mathcal{O}(\epsilon)\;.
\label{adding3}
\eqa
The third term requires a little more thought. Since the 
UV behavior of ${\Pi}^T(P)$ is $1/P^2$, the integrand
${\Pi}_s^0(P){\Pi}^T(P)/P^2$ is 
logarithmically divergent in the ultraviolet.
In order to isolate this divergence, we add and subtract 
${\Pi}^T_{\rm UV}(P)$ 
from ${\Pi}_s^0(P){\Pi}^T(P)/P^2$. Thus the third sum-integral in 
Eq.~(\ref{deco2}) becomes
\bqa\nonumber
I_3&=&2
\sumint_P{1\over P^2}{\Pi}_s^0(P){\Pi}^T(P) \\ 
&=&2\sumint_P^{\prime}{1\over P^2}{\Pi}_s^0(P)
[{\Pi}^T(P)-{\Pi}^T_{\rm UV}(P)]
+2\sumint_P^{\prime}{1\over P^2}{\Pi}_s^0(P){\Pi}^T_{\rm UV}(P)
+2T\int_p{1\over p^2}{\Pi}_s^0(p_0=0,p){\Pi}^T(p_0=0,p)\;,
\label{i3com}
\eqa
where we have isolated the contribution from the
$p_0=0$ term since the contribution to $I_3$ from 
this term is infrared divergent. In order to calculate the first
term in Eq.~(\ref{i3com}), we need ${\Pi}^T_{\rm UV}(P)$
in coordinate space. It is given by the small-$r$ behavior of 
${\Pi}^T(P)$ and reads
\bqa
{\Pi}_{\rm UV}^T(P)
&=&{T\over(4\pi)^2}\int{d^3r\over r^2}e^{i{\bf p}\cdot{\bf r}}
{\bar{r}\over3}
e^{-|p_0|r}\;,
\label{3dptuv}
\eqa 
This yields
\bqa\nonumber
I_3^{\rm a}
&=&2\sumint_P^{\prime}{1\over P^2}{\Pi}_s^0(P)[{\Pi}^T(P)-{\Pi}^T_{\rm UV}(P)]\\
&=&
    \frac{2T^2}{(4\pi)^4} \int d^3r \frac{1}{r^2}
    \left( \coth \bar r - {1\over\bar r} - \frac{\bar r}{3} \right)
    \sum_{p_0 \neq 0} e^{-|p_0|r}
    \int{d^3p\over(2\pi)^3}
    \frac{e^{i {\bf p} \cdot {\bf r}}}{p_0^2 + p^2}
    \left(2 + \log \frac{\mu^2}{p_0^2 + p^2} \right)\;.
    \label{i3a}
\eqa
The integral over 3-momentum can be done analytically.
We write it as
\begin{eqnarray}
    \int {d^3 p \over (2\pi)^3}
    {e^{i {\bf p} \cdot {\bf r}} \over p_0^2 + p^2}
    \left(
        2 + 2 \log{\mu \over 4\pi T}
        + \log{(4\pi T)^2 \over p_0^2 + p^2}
    \right).
\end{eqnarray}
where the first two terms in the parentheses are independent of
$p$, making this part of the integral a simple Fourier transform:
\begin{eqnarray}
    \int {d^3 p \over (2\pi)^3}
    {e^{i {\bf p} \cdot {\bf r}} \over p_0^2 + p^2}
    \left(
        2 + 2 \log{\mu \over 4\pi T}
    \right)
    &=&
    {e^{-|p_0| r} \over 4\pi r}
    \left(
        2 + 2 \log{\mu \over 4\pi T}
    \right).
\end{eqnarray}
Averaging over angles, the last term can be rewritten as
\begin{eqnarray}
    \int {d^3 p \over (2\pi)^3}
    {e^{i {\bf p} \cdot {\bf r}} \over p_0^2 + p^2}
    \log{(4\pi T)^2 \over p_0^2 + p^2}
    &=&
    {1 \over 4\pi^2 i r}
    \int_{-\infty}^\infty dp\,p
    {e^{i p r} \over p_0^2 + p^2}
    \log{(4\pi T)^2 \over p_0^2 + p^2}
\end{eqnarray}
The integrand has a branch cut starting at $p=i|p_0|$
running to $p=i\infty$, and a pole in $p=i|p_0|$.
The contour can be deformed to wrap around the pole
and the branch cut, and taking care to include contributions
from both, one arrives at the result
\begin{eqnarray}
    \int {d^3 p \over (2\pi)^3}
    {e^{i {\bf p} \cdot {\bf r}} \over p_0^2 + p^2}
    \log{(4\pi T)^2 \over p_0^2 + p^2}
    &=&
    {e^{-|p_0| r} \over 4\pi r}
    \left(
        \log{2 \bar r \over |\bar p_0|}
        + \gamma_E
        + e^{2 |p_0| r} {\rm Ei}(-2 |p_0| r)
    \right),
\end{eqnarray}
where $\bar p_0 = p_0/2\pi T = n$ 
and the exponential-integral function ${\rm Ei}(z)$ is defined
as
\begin{eqnarray}
    {\rm Ei}(z) = - \int_{-z}^\infty {dt \; e^{-t} \over t}.
\end{eqnarray}
Thus Eq.~(\ref{i3a}) can be rewritten as
\begin{eqnarray}
    I_3^a
    &=&
    {2 T^2 \over (4 \pi)^4}
    \int d^3 r {1 \over r^2}
    \left(
        \coth \bar r - {1 \over \bar r} - {\bar r \over 3}
    \right)
    \sum_{p_0 \neq 0} {e^{-2 |p_0| r} \over 4\pi r}
    \nonumber \\ && \times
    \left(
        2 + \gamma_E + 2 \log{\mu \over 4\pi T}
        + \log{2 \bar r \over |\bar p_0|}
        + e^{2 |p_0| r} {\rm Ei}(-2 |p_0| r)
    \right).
\label{newia3}
\end{eqnarray}
The first three terms in the last parentheses are independent
of $r$ and $p_0$, and for these terms, the integral over $r$
and the sum over Matsubara modes can be evaluated analytically.
In particular, we are able to find the coefficient of $\log \mu$.
This is fortunate, because it allows us to check the concistency
of our final result for the free energy.
Let
\begin{eqnarray}
    \xi
    &\equiv&
    {2 T^2 \over (4 \pi)^4}
    \int d^3 r {1 \over r^2}
    \left(
        \coth \bar r - {1 \over \bar r} - {\bar r \over 3}
    \right)
    \sum_{p_0 \neq 0} {e^{-2 |p_0| r} \over 4\pi r}.
\end{eqnarray}
Integrating over angles and summing over Matsubara frequencies yields
\begin{eqnarray}
    \xi
    &=&
    {2 T^2 \over (4 \pi)^4}
    \int_0^\infty {d \bar r \over \bar r}
    \left(
        \coth \bar r - {1 \over \bar r} - {\bar r \over 3}
    \right)
    {2 \over e^{2 \bar r} - 1}
    \nonumber \\
    &=&
    {4 T^2 \over (4 \pi)^4}
    \int_0^\infty {d \bar r \over \bar r}
    \left(
        {2 \over e^{2 \bar r} - 1} + 1
        - {1 \over \bar r} - {\bar r \over 3}
    \right)
    {1 \over e^{2 \bar r} - 1}
\end{eqnarray}
The integral above is finite, but the individual terms are
divergent for small $\bar r$. We therefore regulate them by
multiplying by an extra factor $(2\bar r)^\alpha$ and taking
the limit $\alpha \to 0$ in the end.
The basic integrals we need are
\begin{eqnarray}
    \int_0^\infty {dt \; t^x \over e^t - 1}
    &=&
    \Gamma(x+1) \zeta(x+1),
    \\
    \int_0^\infty {dt \; t^x \over (e^t - 1)^2}
    &=&
    \Gamma(x+1) \; [\zeta(x) - \zeta(x+1)].
\end{eqnarray}
This yields
\begin{eqnarray}
    \xi
    &=&
    {4 T^2 \over (4 \pi)^4}
    \left[
        2 \Gamma(\alpha) [\zeta(\alpha-1) - \zeta(\alpha)]
        + \Gamma(\alpha) \zeta(\alpha)
        - 2 \Gamma(\alpha-1) \zeta(\alpha-1)
        - {1 \over 6} \Gamma(\alpha+1) \zeta(\alpha+1)
    \right].
\end{eqnarray}
The limit $\alpha \to 0$ is regular, and we obtain
\begin{eqnarray}
    \xi
    &=&
    - {2 T^2 \over 3 (4 \pi)^4}
    \left(
        1 + \gamma_E - 3 \log(2\pi)
        + 2 {\zeta'(-1) \over \zeta(-1)}
    \right).
\end{eqnarray}
The remaining integral over the coordinate $r$ as well as the Matsubara sum
in Eq.~(\ref{newia3}) must be done numerically. Eq.~(\ref{newia3})
can then be written as  
\bqa
I_3^{\rm a}
&=&
-\frac{2T^2}{3(4\pi)^4} \left[
    \left(2 + \gamma_E + 2\log\frac{\mu}{4\pi T}\right)
    \left(1 + \gamma_E - 3\log(2\pi) + 2\frac{\zeta'(-1)}{\zeta(-1)}\right)
    + C
    \right]\;,
\label{adding4}
\eqa
where the numerical constant $C$ is
\bqa
C&=&
-\frac{3}{4\pi} \int \frac{d^3 r}{r^3}
\left(\coth\bar r - \frac{1}{\bar r} - \frac{\bar r}{3}\right)
\sum_{p_0 \neq 0}
\left(e^{-2 |p_0| r} \log \frac{2 \bar r}{|\bar p_0|} + \mathrm{Ei}(-2 |p_0| r)\right)
= 0.0034814\;.
\eqa
The subtraction term in Eq.~(\ref{i3a}) 
can be calculated with dimensional regularization and
reads
\bqa\nonumber
I_3^{\rm b}&=&2\sumint_P{1\over P^2}{\Pi}_s^0(P){\Pi}^T_{\rm UV}(P)
\\
&=& \nonumber
{4\over(4\pi)^2}
\sumint_Q{1\over Q^2}
\sumint_P{1\over P^4}
\left\{\left({e^{\gamma_E}\mu^2\over P^2}\right)^{\epsilon}
{\Gamma(\epsilon)\Gamma^2(1-\epsilon)\over\Gamma(2-2\epsilon)}
-{1\over\epsilon}\right\}
\\ \nonumber
&=&-{T^2\over6(4\pi)^4}\left[
{1\over\epsilon^2}
+\left(
2\log{\mu\over4\pi T}+2{\zeta^{\prime}(-1)\over\zeta(-1)}
+1\right){1\over\epsilon}
-2\log^2{\mu\over4\pi T}-2\log{\mu\over4\pi T}
\left(1+4\gamma_E-2{\zeta^{\prime}(-1)\over\zeta(-1)}\right)
\right. \\ &&\left.
+2{\zeta^{\prime}(-1)\over\zeta(-1)}
+2{\zeta^{\prime\prime}(-1)\over\zeta(-1)}
-1-{\pi^2 \over 12}-4\gamma_E+8\gamma_1
\right]\;.
\label{i3b}
\eqa
The last term in Eq.~(\ref{i3com}) is
\bqa\nonumber
I_3^{\rm c}&=&
2T\int_p{1\over p^2}{\Pi}^0_s(p_0=0,p){\Pi}^T(p_0=0,p)\\ 
&=&
2T\int_p{1\over p^2}{\Pi}^0_s(p_0=0,p)
\left[{\Pi}(p_0=0,p)-{\Pi}^0(p_0=0,p)
\right]\;,
\label{lasti3}
\eqa
The second term vanishes in dimensional regularization since there
is no mass scale in the integral, i.e.
\bqa
2T\int_p{1\over p^2}{\Pi}_s^0(p_0=0,p){\Pi}^0(p_0=0,p)&=&0\;.
\eqa
In order to evaluate the first term in Eq.~(\ref{lasti3}), we
must calculate ${\Pi}(p_0=0,p)$. Using Feynman parameters,
we obtain
\bqa\nonumber
{\Pi}(p_0=0,p)&=&\sumint_Q{1\over Q^2({\bf p}+Q)^2}
\\ 
&=&T\left({e^{\gamma_E}\mu^2\over4\pi}\right)^{\epsilon} 
{\Gamma(1/2+\epsilon)\over(4\pi)^{3/2-\epsilon}}
\sum_{q_0}\int_0^1{dx\over[x(1-x)p^2+q_0^2]^{1/2+\epsilon}}\;.
\label{qsum}
\eqa
Inserting the expression for ${\Pi}_s^0(p_0=0,p)$ 
and ${\Pi}(p_0=0,p)$, we obtain
\bqa
I_3^{\rm c}&=&{2T^2\over(4\pi)^2}\left({e^{\gamma_E}\mu^2\over4\pi}\right)^{\epsilon} 
{\Gamma(1/2+\epsilon)\over(4\pi)^{3/2-\epsilon}}
\int_p{1\over p^2}
\left[
\left({e^{\gamma_E}\mu^2\over p^2}\right)^{\epsilon}
{\Gamma(\epsilon)\Gamma^2(1-\epsilon)\over\Gamma(2-2\epsilon)}
-{1\over\epsilon}
\right]
\sum_{q_0}\int_0^1{dx\over[x(1-x)p^2+q_0^2]^{1/2+\epsilon}}\;.
\eqa

\begin{eqnarray}
    I_3^{\rm c}
    & = &
        \frac{2T^2}{(4\pi)^4}
        \frac{(e^\gamma \mu^2)^{2\epsilon}}{2\pi}
        \frac{\Gamma(\tfrac{1}{2}+\epsilon)}{\Gamma(\tfrac{3}{2}-\epsilon)}
        \Bigg[
            (e^\gamma \mu^2)^\epsilon
            \frac{\Gamma(\epsilon) \Gamma^2(1-\epsilon)}{\Gamma(2-2\epsilon)}
            \int_0^\infty dp \; \frac{p^{-4\epsilon}}{(p^2+1)^{1/2+\epsilon}}
            \int_0^1 dx \; [x(1-x)]^{-\tfrac{1}{2}+2\epsilon}
            \sum_{q_0}' \frac{1}{|q_0|^{6\epsilon}}
    \nonumber \\
    &&
            -
            \frac{1}{\epsilon}
            \int_0^\infty dp \; \frac{p^{-2\epsilon}}{(p^2+1)^{1/2+\epsilon}}
            \int_0^1 dx \; [x(1-x)]^{-\tfrac{1}{2}+\epsilon}
            \sum_{q_0}' \frac{1}{|q_0|^{4\epsilon}}
        \Bigg]
    \nonumber \\
    & = &
        \frac{2T^2}{(4\pi)^4}
        \left(\frac{e^\gamma \mu^2}{4\pi^2 T^2}\right)^{2\epsilon}
        \frac{\Gamma(\tfrac{1}{2}+\epsilon)}{\Gamma(\tfrac{3}{2}-\epsilon)}
        \Bigg[
            \left(\frac{e^\gamma \mu^2}{4\pi^2 T^2}\right)^\epsilon
            \frac{1}{2^{1+4\epsilon} \sqrt{\pi}}
            \frac{\Gamma(\epsilon) \Gamma^2(1-\epsilon) \Gamma(\tfrac{1}{2}-2\epsilon)
                  \Gamma(3\epsilon) \Gamma(\tfrac{1}{2}+2\epsilon)}
                 {\Gamma(2-2\epsilon) \Gamma(\tfrac{1}{2}+\epsilon) \Gamma(1+2\epsilon)}
            \zeta(6\epsilon)
    \nonumber \\
    &&
            -
            \frac{1}{4\pi\epsilon}
            \frac{\Gamma(\tfrac{1}{2}-\epsilon) \Gamma(\epsilon) \Gamma(\tfrac{1}{2}+\epsilon)}
                 {\Gamma(1+\epsilon)}
            \zeta(4\epsilon)
        \Bigg]\;,
\label{3cc}
\end{eqnarray}
where the prime indicates that we have omitted the $p_0=0$ mode from the sum.
Expanding Eq.~(\ref{3cc}) in powers of $\epsilon$, we obtain
\begin{eqnarray}
    I_3^{\rm c}
    & = &
        \frac{T^2}{6(4\pi)^4}
        \bigg[
            \frac{1}{\epsilon^2}
            - \frac{2}{\epsilon}
            - 12 - \frac{11\pi^2}{3} - 24\log(2\pi) - 12 \log^2(2\pi) - 24\log\frac{\mu}{4\pi T}
    \nonumber \\
    &&
            - 12\log^2\frac{\mu}{4\pi T} - 24\log(2\pi) \log\frac{\mu}{4\pi T}
            + 12\gamma_E^2 + 24\gamma_1
        \bigg]
        + \mathcal{O}(\epsilon)
\label{i3cc}
\end{eqnarray}    

The last term in Eq.~(\ref{deco2}) is
\bqa
I_4&=&
\sumint_P{1\over P^2}[{\Pi}^T(P)]^2\;.
\label{last}
\eqa
Since the UV-behavior of ${\Pi}^T(P)$ is $1/P^2$, the sum-integral 
in Eq.~(\ref{last}) is UV finite.
However, ${\Pi}_T(P)$ has a logarithmic infrared divergence for the 
$p_0=0$ mode. This implies that the sum-integral $I_4$ 
has linear and logarithmic
IR divergences. The linear divergence is set to zero in dimensional 
regularization while the logarithmic is not.
In order to isolate these divergences, we rewrite the sum-integral as
\bqa
I_4&=&\sumint_P^{\prime}{1\over P^2}[{\Pi}^T(P)]^2+
T\int_p{1\over p^2}[{\Pi}^T(p_0=0,p)]^2\;,
\label{split}
\eqa
where the prime indicates that we have omitted the $p_0=0$ mode from the sum.
The primed 
sum-integral in Eq.~(\ref{split}) is finite both in the ultraviolet
and in the infrared. Using the 3-dimensional representation of the
$\Pi_T(P)$, Eq.~(\ref{3dpt}), the first term in Eq.~(\ref{split})
can be written as 
\bqa\nonumber
I_4^{\rm a}&=&
\sumint_P^{\prime}{1\over P^2}[{\Pi}^T(P)]^2
\\  
&=&{T^3\over(4\pi)^4}\sum_{p_0}^{\prime}
\int{d^3p\over(2\pi)^3}\int{d^3r\over r^2}
{d^3r^{\prime}\over(r^{\prime})^2}
{1\over p_0^2+p^2}
\left(\coth\bar{r}-{1\over\bar{r}}\right)
\left(\coth\bar{r}^{\prime}-{1\over\bar{r}^{\prime}}\right)
e^{i{\bf p}\cdot({\bf r}+{\bf r}^{\prime})}e^{-|p_0|(r+r^{\prime})}
\eqa
The integral over three-momentum $p$ corresponds to a Fourier transform
of a massive propagator and so gives rise to a Yukawa potential.
The sum over nonzero Matsubara frequencies can also be done analytically
and we obtain
\begin{eqnarray}
    I_4^{\rm a}
    & = &
        \frac{2 T^3}{(4\pi)^5}
        \int \frac{d^3 r}{r^2} \frac{d^3 r'}{(r')^2} \frac{1}{|\mathbf r + \mathbf r'|}
        \left( \coth \bar r - \frac{1}{\bar r} \right)
        \left( \coth \bar r' - \frac{1}{\bar r'} \right)
    {1\over e^{\bar{r} + \bar{r}^{\prime} + |\bar{\bf r} + \bar{\bf r}^{\prime}|} - 1}
\eqa
Averaging over angles, one finds
\bqa
I_4^{\rm a}    & = &
        \frac{2 T^2}{(4\pi)^4}
        \int_0^\infty \frac{d \bar r \; d \bar r'}{\bar r \bar r'}
        \left( \coth \bar r - \frac{1}{\bar r} \right)
        \left( \coth \bar r^{\prime} - \frac{1}{\bar r'} \right)
    \nonumber \\
    &&
        \qquad\qquad\times\left[
            \log \left( e^{2(\bar r + \bar r')} - 1 \right)
            - \log \left( e^{\bar r + \bar r' + |\bar r - \bar r'|} - 1 \right)
            + |\bar r - \bar r'| - \bar r - \bar r'
            \right]
\end{eqnarray}    
The remaining integrals over $\bar{r}$ and $\bar{r}^{\prime}$ must be done
numerically and we obtain
\bqa
I_4^{\rm a}&=&{T^2\over(4\pi)^4}\left[0.0587392 \right]\;.
\label{i4anum}
\eqa
The second term in Eq.~(\ref{split}) is rewritten as 
\bqa\nonumber
I_4^{\rm b}&=&T\int_p{1\over p^2}[{\Pi}^T(p_0=0,p)]^2
\\ &=& 
T\int_p{1\over p^2}\left\{\left[{\Pi}^T(p_0=0,p)
-\Pi^T_{\rm IR}(p)\right]^2
+2{\Pi}^T(p_0=0,p){\Pi}_{\rm IR}^T(p)
-[{\Pi}_{\rm IR}^T(p)]^2\right\}\;,
\label{rewrite}
\eqa
where ${\Pi}_{\rm IR}(p)$ 
is given by 
the $q_0=0$ term in Eq.~(\ref{qsum}):
\bqa\nonumber
{\Pi}^T_{\rm IR}(p)
&=&T\int_q{1\over q^2({\bf p}+{\bf q})^2}
\\
&=&T\left({e^{\gamma_E}\mu^2\over4\pi}\right)^{\epsilon}
{4^{\epsilon}\sqrt{\pi}
\over(4\pi)^{3/2-\epsilon}}
{\Gamma(1/2+\epsilon)
\Gamma(1/2-\epsilon)\over\Gamma(1-\epsilon)}
p^{-1-2\epsilon}\;.
\eqa
The first integral in Eq.\ \eqref{rewrite} is now
well behaved in both the ultraviolet and the infrared. It can be evaluated
numerically using the representation of $\Pi^T(p_0=0,p)$ in three dimensions.
The subtracted terms are infrared divergent and are calculated with
dimensional regularization.
The first integral can be calculated directly in three dimensions. 
In this case, ${\Pi}^T_{\rm IR}(p)$ reduces to
\bqa
{\Pi}^T_{\rm IR}(p)&=&{T\over8p}\;.
\label{ir}
\eqa
Using the three-dimensional representation~(\ref{3dpt}) for ${\Pi}^T(P)$
with $p_0=0$ and Eq.~(\ref{ir}), we get
\begin{widetext}
\bqa\nonumber
I_4^{\rm b1}&=&
T\int_p{1\over p^2}\left[{\Pi}^T(p_0=0,p)
-{\Pi}_{\rm IR}^T(p)\right]^2
\\\nonumber
&=&
T^3\int_p{1\over p^2}
\left[{1\over(4\pi)^4}\int{d^3r\over r^2}
{d^3r^{\prime}\over(r^{\prime})^2}
e^{i{\bf p}\cdot ({\bf r}+{\bf r}^{\prime})}\left(\coth{r}-{1\over r}\right)
\left(\coth{\bar{r}}^{\prime}-{1\over\bar{r}^{\prime}}\right)
\right.\\&&\left.
-{1\over4(4\pi)^2p}\int{d^3r\over r^2}
e^{i{\bf p}\cdot {\bf r}}
\left(\coth{\bar{r}}-{1\over\bar{r}}\right)
+{1\over64p^2}
\right]\;.
\eqa
\end{widetext}
The averages over the angles between ${\bf p}$ and ${\bf r}$, 
and between ${\bf p}$ and ${\bf r}^{\prime}$ 
can be done analytically and we obtain
\bqa\nonumber
I_4^{\rm b1}&=&T^3\int_p{1\over p^2}
\left[{1\over(4\pi)^2}\int_0^{\infty}
{dr}
\int_0^{\infty}{dr^{\prime}}
{\sin pr\over pr}
{\sin pr^{\prime}\over pr^{\prime}}
\left(\coth{r}-{1\over r}\right)
\left(\coth{\bar{r}}^{\prime}-{1\over\bar{r}^{\prime}}\right)
\right.\\&&\left.
-{1\over4(4\pi)p}\int_0^{\infty}{dr}{\sin pr\over pr}
\left(\coth{\bar{r}}-{1\over\bar{r}}\right)
+{1\over64p^2}
\right]\;.
\eqa
The integrals 
over $r$, $r^{\prime}$, and $p$ must be done numerically. The result
is 
\bqa
I_4^{\rm b1}&=&{T^2\over(4\pi)^4}[9.5763]\;.
\label{b1sub}
\eqa

The first subtraction term in Eq.~(\ref{rewrite}) is
\bqa\nonumber
I_4^{\rm b2}&=&2T
\int_p{1\over p^2}
{\Pi}_T(p_0=0,p){\Pi}^T_{\rm IR}(p)
\\ &=& \nonumber
2T\int_p{1\over p^2}\left[{\Pi}(p_0=0,p)-{\Pi}^0(p_0=0,p)\right]
{\Pi}^T_{\rm IR}(p)
\\ &=& 
2T\int_p{1\over p^2}{\Pi}(p_0=0,p){\Pi}^T_{\rm IR}(p)\;,
\label{b1sub1}
\eqa
where we have used that the second term vanishes in dimensional regularization:
This term is logarithmically divergent both in the infrared and in the
ultraviolet. If we use the same scale for the regularization of ultraviolet
and and infrared divergences, the integral vanishes~\cite{ericnieto1}.

Inserting the expressions for 
${\Pi}^0(p_0=0,p)$ and ${\Pi}^T_{\rm IR}(p)$ into Eq.~(\ref{b1sub1}),
we obtain
\begin{widetext}
\begin{eqnarray}
    I_4^{\rm b2}
    & = &
        \frac{T^3}{(4\pi)^{4-3\epsilon}}
        \left(\frac{e^{\gamma_E} \mu^2}{4\pi}\right)^{3\epsilon}
        2^{1+2\epsilon}
        \frac{\Gamma^2(\tfrac{1}{2}+\epsilon) \Gamma(\tfrac{1}{2}-\epsilon)}
             {\Gamma(\tfrac{3}{2}-\epsilon) \Gamma(1-\epsilon)}
        \int_0^\infty dp  \int_0^1 dx \sum_{q_0}
        \frac{p^{-1-4\epsilon}}{[x(1-x) p^2 + q_0^2]^{\frac{1}{2}+\epsilon}}
    \nonumber \\
    & = &
        \frac{T^3}{(4\pi)^{4-3\epsilon}}
        \left(\frac{e^{\gamma_E} \mu^2}{4\pi}\right)^{3\epsilon}
        2^{1+2\epsilon}
        \frac{\Gamma^2(\tfrac{1}{2}+\epsilon) \Gamma(\tfrac{1}{2}-\epsilon)}
             {\Gamma(\tfrac{3}{2}-\epsilon) \Gamma(1-\epsilon)}
        \int_0^\infty dp \; \frac{p^{-1-4\epsilon}}{(p^2+1)^{\tfrac{1}{2}+\epsilon}}
        \int_0^1 dx \; [x(1-x)]^{2\epsilon}
        \sum_{q_0}^{\prime}  \frac{1}{|q_0|^{1+6\epsilon}}
    \nonumber \\
    & = &
        \frac{T^2}{(4\pi)^4}
        \left(\frac{e^{\gamma_E} \mu^2}{4\pi^2 T^2}\right)^{3\epsilon}
        \frac{4^\epsilon}{\pi}
        \frac{\Gamma(\tfrac{1}{2}+\epsilon) \Gamma(\tfrac{1}{2}-\epsilon)
              \Gamma(-2\epsilon) \Gamma(\tfrac{1}{2}+3\epsilon) \Gamma^2(1+2\epsilon)}
             {\Gamma(\tfrac{3}{2}-\epsilon) \Gamma(1-\epsilon) \Gamma(2+4\epsilon)}
        \zeta(1+6\epsilon)\;.
\label{puha}        
\end{eqnarray}
The prime on the sum in the second line indicates that we have excluded
the zero mode, $q_0=0$, from the sum. 
This mode gives rise to an integral
that is linearly divergent in the infrared. Since there is no mass scale
in this integral, it vanishes. Note also that the integral over $p$
is logarithmically divergent in the infrared and this divergence is {\it not}
set to zero in dimensional regularization~\cite{4loopfinns}.
Expanding Eq.~(\ref{puha}) in powers of $\epsilon$, we obtain
\begin{eqnarray}
    I_4^{\rm b2}
    & = &
        -\frac{T^2}{6(4\pi)^4} \bigg[
            \frac{1}{\epsilon^2}+
\left(6\log\frac{\mu}{4\pi T} + 6\gamma_E -2\right)\frac{1}{\epsilon} 
            + 12 + {25\over12}\pi^2 - 12\log\frac{\mu}{4\pi T} + 
18\log^2\frac{\mu}{4\pi T}
    \nonumber \\
    && \qquad \qquad
            + 36\gamma_E \log\frac{\mu}{4\pi T} - 12\gamma_E -36\gamma_1
        \bigg] + \mathcal{O}(\epsilon)\;.
\label{i4b2}
\end{eqnarray}    
\end{widetext}

Finally, we consider the last subtraction term
in Eq.~(\ref{rewrite}). Since 
${\Pi}_{\rm IR}^T(p_0=0,p)$
goes like $1/p$ for small $p$, the integrand has a linear infrared divergence.
This divergence is set to zero in dimensional regularization. In fact,
since there is no mass scale in the integral, it vanishes:
\bqa
T\int_p{1\over p^2}[{\Pi}_{\rm IR}^T(p)]^2&=&0\;.
\eqa

Adding Eqs.\ (\ref{adding2}), (\ref{adding3}), (\ref{adding4}), (\ref{i3b}),
(\ref{i3cc}), (\ref{i4anum}), (\ref{b1sub}), 
and~(\ref{i4b2}), we can write $I$ in the following form:
\begin{widetext}

\begin{eqnarray}
\nonumber
I
    & = &
        -\frac{T^2}{4(4\pi)^4}\left({\mu\over4\pi T}\right)^{6\epsilon}
 \bigg\{
            \frac{1}{\epsilon^2}
            + \frac{1}{\epsilon} \left[
\frac{4}{3}+2\frac{\zeta'(-1)}{\zeta(-1)} + 4\gamma_E 
                \right]
            + \frac{1}{3} \bigg[46 - 8\gamma_E
- 16\gamma_E^2 
- 104\gamma_1 - 24\gamma_E \log(2\pi)
\\ &&
                + 24\log^2(2\pi) + 
\frac{45\pi^2}{4} 
+ 24\frac{\zeta'(-1)}{\zeta(-1)}
+ 2\frac{\zeta''(-1)}{\zeta(-1)}
                + 16\gamma_E \frac{\zeta'(-1)}{\zeta(-1)}
                \bigg]
-38.5309
+{\cal O}(\epsilon)            \bigg\}\;.   
\end{eqnarray}    

\end{widetext}

\appendix
\renewcommand{\theequation}{\thesection.\arabic{equation}}

\end{document}